%% file: main.tex
\documentclass[manuscript, nonacm]{acmart}

\renewcommand\footnotetextcopyrightpermission[1]{}
\usepackage{booktabs}   
\usepackage{subcaption}  
\usepackage{graphicx}
\usepackage{hyperref}
\usepackage{cleveref}

\title{Software as Content: Dynamic Applications as the Human-Agent Interaction Layer}

\author{Mulong Xie$^*$}
\thanks{$^*$First \& Corresponding author}
\email{mulong@mulongxie.me}

\author{Yang Xie}
\email{xieyang@fellou.ai}

\input{sections/0.abstract}


\begin{teaserfigure}
  \includegraphics[width=\textwidth]{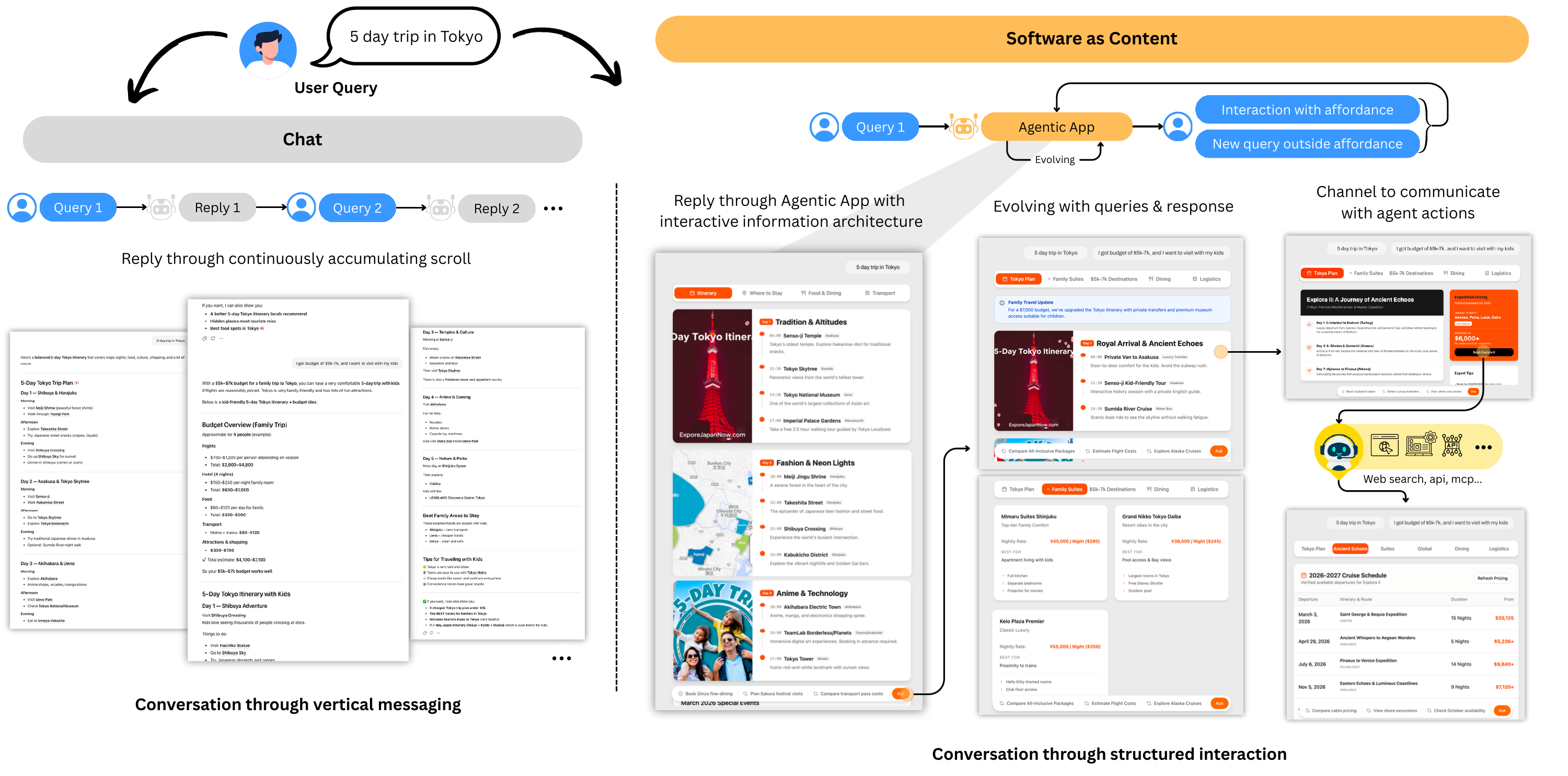}
  \caption{Given the same user query, chat returns a continuously 
  accumulating conversation wall (left); Software as Content generates 
  an evolving agentic application that serves as a bidirectional 
  interaction layer between human and agent (right).}
  \label{fig:teaser}
\end{teaserfigure}

\begin{document}
\maketitle

\section{Introduction}
\label{sec:intro}
\input{sections/1.intro}

\section{Related Work}
\label{sec:related}
\input{sections/2.related-work/2.1.Human-Agent-Interaction-and-the-Chat-Paradigm}
\input{sections/2.related-work/2.2.Generative-User-Interfaces}
\input{sections/2.related-work/2.3.Emerging-Agent-UI-Protocols}
\input{sections/2.related-work/2.4.Interaction-Design-and-Cognitive-Foundations}
\input{sections/2.related-work/2.5.AI-Powered-Software-Generation}

\section{Software as Content: A Conceptual Framework}
\label{sec:framework}
\input{sections/3.framework/3.1.The-Human-Agent-Environment-Interaction-Model}
\input{sections/3.framework/3.2.Agentic-Applications-as-a-Content-Modality}
\input{sections/3.framework/3.3.Lifecycle}
\input{sections/3.framework/3.4.A-Taxonomy-of-Agentic-Applications}
\input{sections/3.framework/3.5.Design-Principles-for-SaC-Systems}

\section{System Architecture}
\label{sec:system}
\input{sections/4.system-architecture/4.1.Overview}
\input{sections/4.system-architecture/4.2.App-Generation}
\input{sections/4.system-architecture/4.3.App-Evolution}
\input{sections/4.system-architecture/4.4.Quality-Assurance}
\input{sections/4.system-architecture/4.5.Distribution-and-Consumption}

\section{Evaluation}
\label{sec:evaluation}
\input{sections/5.evaluation/5.1.Methodology}
\input{sections/5.evaluation/5.2.Scenario-Walkthroughs}
\input{sections/5.evaluation/5.3.Boundaries-of-the-SaC-Scope}

\section{Discussion}
\label{sec:discussion}
\input{sections/6.discussion/6.1.Rethinking-the-Role-of-the-Interface-in-Agent-Systems}
\input{sections/6.discussion/6.2.Limitation}
\input{sections/6.discussion/6.3.Future-Work}

\section{Conclusion}
\label{sec:conclusion}
\input{sections/7.conclusion}


\bibliographystyle{ACM-Reference-Format}
\bibliography{ref}





\end{document}

%% file: sections/0.abstract.tex
\begin{abstract}
Chat-based natural language interfaces have emerged as the dominant paradigm for human–agent interaction, yet they fundamentally constrain engagement with structured information and complex tasks. We identify three inherent limitations: the mismatch between structured data and linear text, the high entropy of unconstrained natural language input, and the lack of persistent, evolving interaction state.
We introduce Software as Content (SaC), a paradigm in which dynamically generated agentic applications serve as the primary medium of human–agent interaction. Rather than communicating through sequential text exchange, this medium renders task-specific interfaces that present structured information and expose actionable affordances—such as filters, selectors, and triggers—through which users iteratively guide agent behavior without relying solely on language.
Crucially, these interfaces persist and evolve across interaction cycles, transforming from transient responses into a shared, stateful interaction layer. Through this process, the system progressively converges toward personalized, task-specific software. We operationalize this paradigm through a multi-level rendering strategy supporting incremental refinement, structural extension, and full interface reconfiguration.
We formalize SaC through a human–agent–environment interaction model and derive design principles for generating and evolving agentic applications. We implement a working system and evaluate it across representative tasks of selection, exploration, and execution, demonstrating technical viability and expressive range across diverse task types. We further identify boundary conditions under which natural language remains preferable.
By reframing interfaces as dynamically generated and distributable software artifacts, SaC opens a new design space for human–AI interaction, positioning dynamic software as a concrete and tractable research object.
\end{abstract}

%% file: sections/1.intro.tex

The capabilities of AI agents have advanced at a remarkable pace.
Modern large language model (LLM) based agents can browse the web, execute code, invoke APIs, query databases, and orchestrate multi-step workflows that would have been unthinkable just two years ago~\cite{yao2023react, schick2023toolformer, anthropic2024computeruse}.
Yet despite this rapid progress in what agents can do, the dominant paradigm for how agents communicate with humans has remained largely unchanged: the chat interface.
The chat paradigm---a sequential exchange of natural language messages between user and agent---is fundamentally a \emph{message-passing} interaction model: two entities communicate by dispatching discrete, self-contained messages across a serial text channel.
As agent capabilities grow, this architecture creates a structural misalignment between modern agent capabilities and the medium through which they are expressed, and it exposes three systemic constraints that no amount of prompt engineering or response formatting can resolve.

\textbf{Representation mismatch.}
Chat is a linear channel.
Both directions of communication---agent to human and human to agent---must be serialized into sequential text.
Yet the information that agents operate over is increasingly structured: comparison tables, spatial relationships, hierarchical categories, or even a series of actions.
Flattening these into prose discards the structure that makes them readable and intuitive.
A multi-attribute comparison across rental car providers becomes a wall of paragraphs; a set of geographically distributed options loses its spatial relationships entirely.
The same problem applies in reverse: a user's intent is often inherently parametric (``cheaper,'' ``closer to the airport,'' ``compare these two'') but must be wrapped in a full natural language sentence to traverse the chat channel.

\textbf{Interaction entropy.}
At each turn, the user faces an empty text box and the full expanse of natural language.
What can the agent do next?
What follow-up questions are meaningful given the current context?
What parameters can be adjusted?
These are invisible---the user must independently infer the productive next step and formulate it as a prompt.
This is a high-entropy interaction: the space of possible inputs vastly exceeds the space of useful inputs, and the entire burden of navigating this gap falls on the user.
Furthermore, conversation context still accumulates implicitly in the chat log---a lossy, append-only medium that neither party can efficiently index or restructure.

\textbf{Ephemeral interaction state.}
Chat treats each exchange as a discrete transaction.
Any intermediate representation that emerges during the conversation---a table, a set of filtered results, or even a generative UI component---exists only as a segment of scroll history.
The user can only refine a previous result by verbalizing the desired change as a new natural language message.
Recent systems such as Claude Artifacts~\cite{anthropic2024artifacts} and OpenAI's Canvas~\cite{openai2024canvas} partially address persistence by maintaining a generated artifact alongside the conversation, allowing iterative refinement.
However, the interaction model remains unchanged: the user inspects the artifact, returns to the chat input, describes the modification in text, and the agent updates or regenerates the artifact accordingly.
The artifact is a display surface, not an interaction layer---it does not serve as a bidirectional channel through which user actions flow back to the agent.

Recent work on \emph{generative user interfaces} (GenUI) addresses the representation mismatch in the agent-to-human direction.
By producing interactive UI components---tables, charts, maps, cards---rather than plain text, GenUI systems enable richer, more navigable presentations of agent output.
The capability is gaining traction in academia~\cite{leviathan2025genui, chen2025genui} and seeing commercial deployment by major AI companies~\cite{anthropic2024artifacts, openai2024canvas, google2025genui, vercel2023v0}.
However, this approach operates at the level of output representation, not interaction.
GenUI as currently conceived addresses only the first of the three constraints, and only in the agent-to-human direction.
The generated interface remains a terminal artifact---a one-shot rendering produced in response to a query, not a persistent medium through which subsequent interaction flows.
The user's next action is still to type another message into the same unconstrained text box: the interaction entropy is unchanged, and the interface is ephemeral.
GenUI improves the presentation of agent responses but does not alter the underlying interaction architecture.

We propose a fundamental reframing of the human--agent interaction paradigm: the interface is no longer the output of interaction, but its medium.
We introduce \textbf{Software as Content (SaC)}, a paradigm in which dynamically generated interactive applications---which we term \textbf{agentic applications}---serve not merely as agent output, but as the bidirectional interaction layer between humans and AI agents.
We use the word ``content'' deliberately: just as text, images, and video are content generated for human consumption, an agentic application is a piece of software generated on demand and consumed interactively.
The difference is that this content is not passive---it is a live, evolving medium through which human and agent continuously communicate.

In the SaC paradigm, three layers---\emph{human}, \emph{agent}, and \emph{environment} (external information and services)---are connected through the agentic application, which mediates their interaction bidirectionally and directly addresses the three constraints identified above:

\begin{itemize}
    \item \textbf{Agent $\rightarrow$ Human (resolving representation mismatch).}
    The agent constructs a complete \emph{information architecture}---not merely individual generative UI components, but an integrated layout that organizes, prioritizes, and spatially relates information to reduce cognitive load.
    Rather than presenting isolated outputs, the agent determines how information is structured, grouped, and revealed based on task context, retrieved data, and interaction history.
    The result is closer to a purpose-built application than a rendered response.

    \item \textbf{Human $\rightarrow$ Agent (resolving interaction entropy).}
    Instead of composing natural language instructions, users interact through familiar UI affordances---filters, selectors, buttons, sliders---that translate intent into structured directives.
    These affordances expose the agent's actionable space---available operations, adjustable parameters, and meaningful follow-ups---as concrete, discoverable elements.
    This shifts the user's cognitive task from \emph{recall} (``what should I ask next?'') to \emph{recognition} (``which option do I select?'')~\cite{nielsen1994usability}.
    In this model, the agent acts as the backend: each interaction directly triggers agent execution, and users need not always compose prompts---or even be aware of the underlying agent---to drive the system forward.

    \item \textbf{Continuous evolution (resolving ephemeral state).}
    The agentic application is not a static artifact, but evolves across interaction cycles: user actions trigger agent execution, whose results update the interface and expose new affordances.
    Over time, the application progressively specializes to the user's task and preferences, converging toward what we term \emph{personalized software}---a task-specific tool that emerges through interaction rather than being designed a priori.
    The application itself becomes the interaction state: persistent, structured, and directly manipulable by both parties.
\end{itemize}

Importantly, SaC does not eliminate natural language input---it complements it.
No generated interface can anticipate every direction a user's intent may evolve, so SaC retains a natural language channel alongside structured affordances.
The key shift is that structured interaction becomes the primary mode for high-frequency, parameterized operations, while natural language remains an always-available channel for expressing unanticipated or nuanced intent.
We formalize this complementary relationship in \S\ref{sec:agentic-app}.

This paper makes the following contributions:
\begin{enumerate}
    \item We introduce the \textbf{Software as Content (SaC) paradigm} and formalize its underlying interaction model as a human--agent--environment loop mediated by agentic applications, whose state---comprising view, affordance set, and agent context---evolves across interaction cycles through a dual-channel model of structured affordances and natural language (\S\ref{sec:interaction-model}--\S\ref{sec:agentic-app}).

    \item We characterize the \textbf{lifecycle and design space of agentic applications}, including patterns of progressive specialization over interaction cycles and a taxonomy along intent, data, and temporal dimensions (\S\ref{sec:lifecycle}).

    \item We derive a set of \textbf{design principles} for SaC systems, governing the integration of generative interfaces and agentic execution, including interface coherence, capability boundaries, and progressive personalization (\S\ref{sec:principles}).

    \item We present a \textbf{system architecture and reference implementation} that operationalizes the SaC paradigm, demonstrating its technical feasibility across diverse task domains (\S\ref{sec:system}).

    \item We conduct a \textbf{scenario-based evaluation} across selection, exploration, and execution tasks, demonstrating expressive range and technical viability, while identifying boundary conditions under which structured interaction is not the appropriate modality (\S\ref{sec:evaluation}).
\end{enumerate}

More broadly, SaC positions \emph{dynamic software}---generated on demand, evolved through use, and disposable once its purpose is served---as a concrete research object, raising new questions in its design, evaluation, and distribution (\S\ref{sec:discussion}).

%% file: sections/2.related-work/2.1.Human-Agent-Interaction-and-the-Chat-Paradigm.tex

\subsection{Human--Agent Interaction and the Chat Paradigm}
\label{sec:related:agents}

The capabilities of LLM-based agents have advanced at a remarkable pace.
ReAct~\cite{yao2023react} demonstrated that reasoning and acting could be interleaved in a unified paradigm, allowing agents to query external environments mid-reasoning and correct themselves in response to new observations.
Frameworks such as LangChain~\cite{langchain2023} and AutoGen~\cite{wu2023autogen} made it straightforward to compose agents that coordinate across tools and, in AutoGen's case, across multiple agents communicating through structured conversation.
More recently, a parallel line of work has extended the agent's action space to direct GUI interaction: CogAgent~\cite{hong2024cogagent} demonstrated that visual language models could navigate both PC and mobile interfaces from screenshots alone; AppAgent~\cite{zhang2023appagent} enabled autonomous smartphone automation through multimodal visual perception; and frontier systems such as Anthropic's computer use~\cite{anthropic2024computeruse} brought these capabilities to production, allowing agents to operate arbitrary software interfaces through the same pixel-level observations a human would use.
Surveys of this literature confirm the scope of the progress: LLM-based agents now operate across web navigation, code execution, API orchestration, and long-horizon planning~\cite{xi2023rise}.

What has not kept pace with this expansion in capability is a reconsideration of the channel through which agents and humans communicate~\cite{luger2016badpa}.
Across the works above---from tool-calling agents to multi-agent systems to computer use---the human--agent interface follows a single, shared structure: the user expresses intent as natural language text, and the agent responds with natural language text or formatted markdown.
As agent capabilities have expanded from API calls to full desktop control, the medium through which they communicate with the user has remained largely textual-oriented.
This pattern reflects not a deliberate design decision but an inherited assumption: that the chat channel is the appropriate medium for human--agent communication, regardless of what the agent is doing on the user's behalf.
The assumption is so pervasive that it largely goes unremarked.

The most direct evidence of this blind spot comes from work that does take the human--agent interface seriously ~\cite{amershi2019guidelines}.
Zou et al.~\cite{zou2025humanagentsurvey} provide the first comprehensive survey of LLM-based human--agent systems, cataloguing how systems vary along five dimensions: environment, human feedback type, interaction pattern, orchestration, and communication structure.
Their framework is sophisticated, and their analysis of how different feedback types---evaluative, corrective, guidance-based, implicit---shape system behavior is valuable.
Yet the survey's taxonomy implicitly preserves the chat channel throughout: human feedback is treated as a signal flowing into the agent's reasoning, not as an action flowing through a shared interactive medium.
The question of what form the agent's output takes, and whether that form constitutes an interface through which the human can act directly, falls outside the survey's scope.

This gap is not a criticism of that work; it reflects the field's shared framing.
The dominant assumption---that improving human--agent collaboration means improving how agents receive and process human text input---has produced genuine advances ~\cite{yang2020haidesign}.
But it leaves a structural question unasked: whether a textual-oriented channel, as the dominant medium of human--agent communication, is the appropriate architecture for tasks where the information is structured, the decisions are multi-dimensional, and the interaction must evolve across cycles.
SaC addresses this question directly by proposing a different interaction architecture---one in which a dynamically generated application, rather than a text channel, mediates communication between human and agent.

%% file: sections/2.related-work/2.2.Generative-User-Interfaces.tex

\subsection{Generative User Interfaces}
\label{sec:related:genui}

\begin{figure}[t]
  \centering
  \includegraphics[width=\columnwidth]{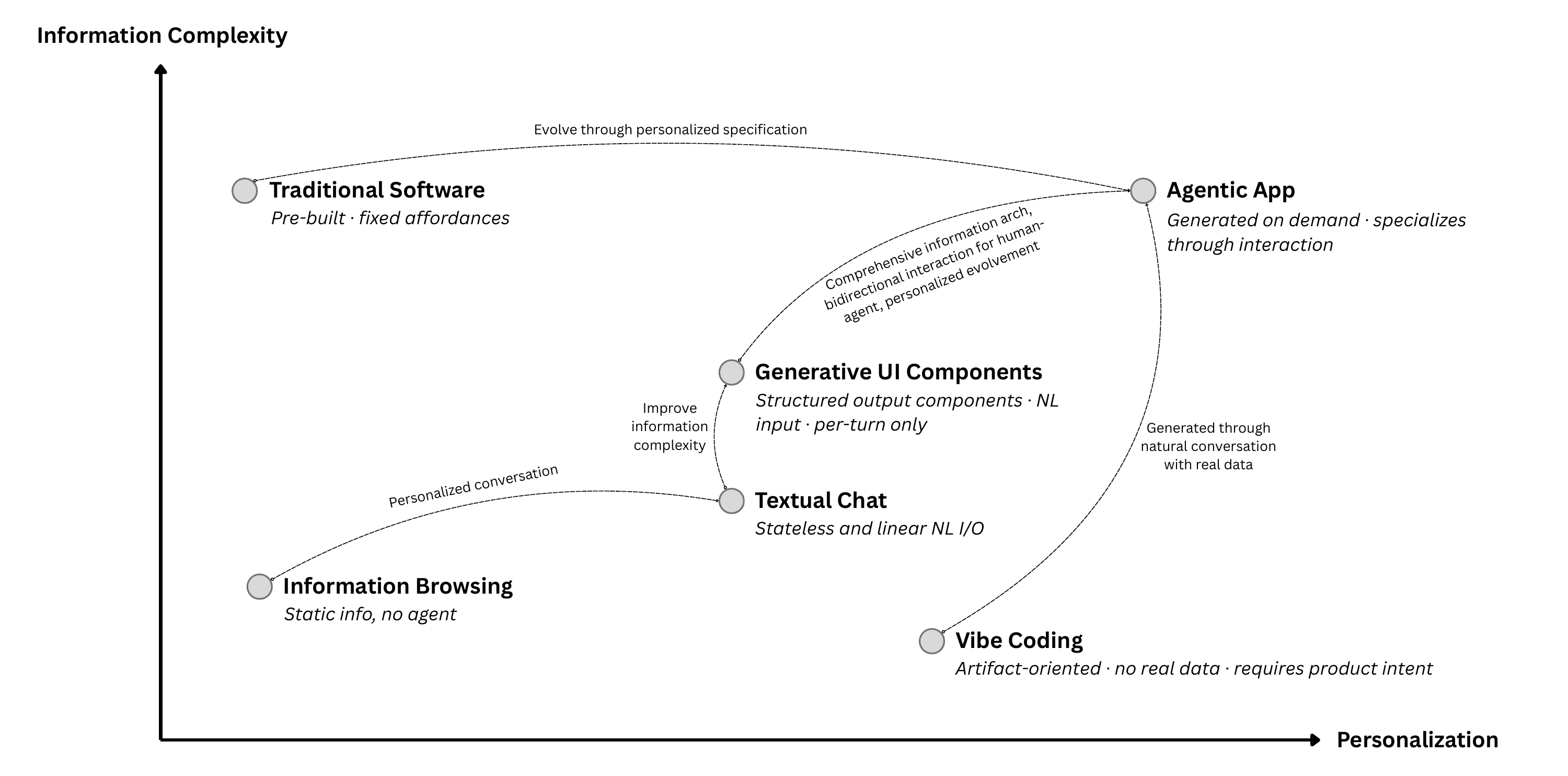}
  \caption{Positioning and comparison of existing interaction paradigms along two dimensions:
  \textit{information complexity}---the degree to which a task requires structured, organized information architecture rather than linear presentation;
  \textit{personalization}---the degree to which the interaction medium adapts to the specific user's task over time.
  Information browsing and textual chat occupy the low end of both axes.
  Generative UI components improve information complexity by producing structured output, but remain stateless and per-turn, offering no bidirectional interaction for human and agent.
  Traditional software achieves high information complexity through pre-built structured interfaces, but personalization is limited to fixed settings.
  Vibe coding achieves high personalization through artifact-oriented construction, but operates on empty containers disconnected from real task data, and requires explicit product intent from the user.
  }
  \label{fig:ic-personalization}
\end{figure}

A parallel line of work has begun to question not agent capability, but agent output format.
The dominant response format for LLM-based systems---markdown rendered in a chat panel---is, on reflection, a choice rather than a necessity: a text model of what the response should contain, rendered as the closest approximation a fixed UI can produce.
This observation motivates the concept of \emph{generative user interfaces} (GenUI): a modality in which the model generates not only content but the entire interactive surface through which that content is presented~\cite{leviathan2025genui, jiang2023graphologue}.
Leviathan et al.~\cite{leviathan2025genui} provide the foundational capability validation: given the right prompts and tools, modern LLMs can generate high-quality custom interactive interfaces for a wide range of prompts, and human raters overwhelmingly prefer the results over markdown output.
The study also releases PAGEN, a dataset of expert-crafted interfaces to support systematic evaluation of future GenUI implementations.
Chen et al.~\cite{chen2025genui} extend this line by proposing a structured framework for generating interfaces in response to multi-turn, information-dense queries.
Their system formalizes interface structure using finite state machines and applies iterative generation-evaluation cycles to refine candidates; a user study across diverse task types confirms substantial advantages in preference, task efficiency, and user satisfaction compared to chat-based baselines.
Taken together, these works establish that GenUI is technically viable and that users respond positively to it---the enabling condition for any subsequent argument about its role in human--agent interaction.

With feasibility established, subsequent work has asked how GenUI integrates into ongoing conversation rather than appearing as a one-shot response.
GenerativeGUI~\cite{hojo2025generativegui} embeds dynamically generated HTML widgets directly within a chat interface, generating a new GUI at each conversation turn rather than presenting a fixed, pre-designed panel.
A user study confirms that the generated GUIs reduce mental demand and task completion time compared to pure text exchange, though the study also finds that GUI generation can reduce effectiveness when the number of available clarification options is constrained---a finding that reveals the limits of treating GUI generation as a drop-in replacement for text.
The Jelly system~\cite{malleableui2025} takes a more structured approach: rather than generating interface code directly, it interposes a declarative intermediate representation---an object-relational schema and dependency graph---between the user's natural language prompt and the rendered UI.
The LLM generates and updates this schema; the UI is then derived from it according to fixed mapping rules.
Users can refine the interface by issuing further prompts or by modifying the schema directly, with changes propagating through to the rendered view.
The target scenario is open-ended information organization, and no persistent agent operates in the loop: the LLM is invoked to update the schema model, not to pursue a task on the user's behalf.
The distinction from SaC is therefore not only structural but motivational: Jelly is a tool for users to build and reshape their own information spaces, whereas SaC is an interaction layer through which a task-executing agent and a user communicate across the lifecycle of a shared task.

The conceptual landscape underlying these systems has recently been formalized.
Lee~\cite{lee2025genuidefinition} synthesizes a working definition of GenUI from a systematic review of 127 publications, 18 expert interviews, and 12 case analyses, identifying five constitutive dimensions: computational co-creation, expanded design space exploration, representation fluidity, contextual adaptation, and synthesis over selection.
The definition positions GenUI as a new mode of interface creation that emphasizes design-time collaboration between human and machine, with users interacting at run time with AI-generated GUI variants.
This framing is precise and useful, and it clarifies the scope of what current GenUI work encompasses.
It also clarifies what it does not encompass: the five dimensions are all properties of how the interface is \emph{generated}, not of how the interface functions as a channel through which an agent receives structured input from the user.
The question of whether the generated interface constitutes a medium of communication---one through which both human and agent act, not merely one through which the agent responds---falls outside the definition's scope.

This gap is structural, not incidental.
Across the GenUI literature, the generated artifact occupies a consistent position: it is the terminal output of an agent turn, a surface that the human reads, interacts with, and may respond to via natural language.
The interface, in all these systems, is an output that the human can read and respond to; it is not a medium through which human and agent jointly act~\cite{beaudouinlafon2000instrumental}.
SaC differs from GenUI not in degree but in the structural role assigned to the generated application.
In SaC, the application is not the response to a user turn; it is the interaction layer that persists across turns, accumulates state, and constitutes the shared surface through which human intent and agent action are mutually expressed.
A single-cycle GenUI interaction---where an agent generates a UI and the human responds via text---is precisely the degenerate case of SaC in which the application's state does not evolve beyond its initial render.
The distinction between GenUI and SaC is therefore the distinction between artifact-as-output and artifact-as-medium.

%% file: sections/2.related-work/2.3.Emerging-Agent-UI-Protocols.tex

\subsection{Emerging Agent--UI Protocols}
\label{sec:related:protocols}

The convergence of industry and open-source efforts around agent-driven interfaces has accelerated substantially since 2025, producing a cluster of complementary protocols that address the infrastructure layer of the problem GenUI research identified.
AG-UI~\cite{agui2025}, developed by CopilotKit, defines an event-based transport protocol that standardizes how agent backends stream execution state---messages, tool calls, state patches, lifecycle signals---to a frontend application, so that developers no longer need to implement custom synchronization logic for each agent framework they integrate.
Google's A2UI~\cite{google2025a2ui} operates at a different layer: rather than governing the transport connection, it specifies a declarative JSON format through which agents describe component trees that client applications render using their own native UI frameworks, avoiding the security risks of transmitting executable code across trust boundaries.
MCP Apps~\cite{mcpapps2026}, a jointly authored extension to the Model Context Protocol from Anthropic and OpenAI, formalizes a third piece: MCP tools can now declare interactive HTML interfaces alongside their text responses, rendered inline within compliant hosts such as Claude and ChatGPT.
The rapid co-emergence of these three protocols---each addressing a distinct layer of the agent-to-UI stack---reflects a broad industry recognition that text-only agent responses are structurally inadequate, and that the infrastructure for richer output modalities must be standardized rather than left to individual implementations.

Taken together, these protocols move the agent's output format from text to structured UI---but leave the interaction model unchanged (as shown in Figure~\ref{fig:ic-personalization}): the user still operates on a separate input channel, and the interface still resets with each new turn.
MCP Apps is the most instructive case: its bidirectional communication allows a rendered UI to invoke tool calls and receive updated data, which is a meaningful step beyond static output.
But this bidirectionality is scoped to a single tool response lifecycle.
The interface does not persist into the next conversation turn; it does not accumulate state across the task; it cannot evolve in response to subsequent agent execution.
Each tool call produces an independent UI instance that exists only within its originating turn, and the next user action still requires composing a new message into the same text channel.
The distinction SaC draws is therefore not between unidirectional and bidirectional protocols---it is between interfaces that exist within turns and an interface that exists across them.
In SaC, the agentic application is not produced by a turn; it is the persistent workspace through which the entire task lifecycle---all its turns, all its state, all its evolution---is conducted, making the full history of human--agent collaboration directly visible and manipulable by both parties at every moment.

%% file: sections/2.related-work/2.4.Interaction-Design-and-Cognitive-Foundations.tex

\subsection{Interaction Design and Cognitive Foundations}
\label{sec:related:cognition}

The shift from command-line to graphical user interfaces in the 1980s is often described as a change in aesthetics, but it is more accurately understood as a change in cognitive architecture.
Norman's analysis of affordances establishes the underlying principle: well-designed objects make their possible actions perceptible, so that users can act without first constructing an internal model of the system's operation~\cite{norman1988design}.
Nielsen's formalization of this principle as a usability heuristic is direct: interfaces should present options rather than require users to recall them, because recognition is cognitively cheaper and more reliable than recall~\cite{nielsen1994usability}.
The GUI did not make computing more attractive; it made the action space visible.
A user navigating a file system through a graphical interface can see what exists, what is selectable, and what operations are available; a user navigating the same system through a command-line interface must hold all of this in memory and express it as syntax.
The cognitive demand is structurally different, and the historical adoption of GUIs over CLIs reflects a broad empirical verdict on which structure better fits human cognitive capacity.

LLM-based chat interfaces, despite their apparent naturalness, reproduce the cognitive structure of the command-line interface at a higher level of abstraction.
The user must hold in working memory the current task state, the steps already completed, the constraints established in prior turns, and the range of actions the agent is capable of taking---none of which are externalized in the interface.
Each turn begins from a blank text field, with no persistent representation of where the task stands or what options are available.
The naturalness of language conceals this demand, but does not reduce it: asking an agent to perform a complex multi-step task through chat requires the user to maintain a cognitive model that the interface itself never displays.
This is the structural source of what we term \emph{interaction entropy}: the accumulating cost of re-expressing, clarifying, and reconstructing shared context across turns, not because the agent lacks capability, but because the channel provides no persistent surface on which that context can be anchored.

Shneiderman's theory of direct manipulation offers a precise account of what is missing~\cite{shneiderman1983direct, hutchins1985direct}.
Direct manipulation rests on three conditions: a continuous representation of the objects of interest, physical actions on those representations rather than complex syntax, and rapid incremental feedback that makes the effects of actions immediately visible.
These conditions explain, at a mechanistic level, why GUIs succeeded where CLIs struggled: the objects being operated on are persistently present, and the user's actions have visible, reversible effects on those objects.
LLM chat satisfies none of these conditions---the task state is not represented, user actions are syntactic, and the effect of a given message on the agent's understanding is opaque until the next response arrives.
SaC can be understood as the application of direct manipulation principles to agent-mediated task work: the agentic application is a continuous representation of the task state, user interactions are direct actions on that representation rather than descriptions of desired state changes, and the application's evolution across cycles provides the incremental feedback that makes the human-agent collaboration legible.
The insight is not that chat is a bad interface in general; it is that chat is a bad interface for tasks whose objects need to be seen, acted upon, and evolved.
SaC is not a replacement for chat but a complement to it: for lightweight, ephemeral exchanges where the task is simple and the response terminal, a text turn is the appropriate medium.
SaC and chat coexist within the same interaction architecture---the same agent can respond in text when a sentence suffices and instantiate an application when the task demands structure, persistence, and direct manipulation.
The two modalities are not in competition; they occupy different regions of the same task space.

%% file: sections/2.related-work/2.5.AI-Powered-Software-Generation.tex

\subsection{AI-Powered Software Generation}
\label{sec:related:softwaregen}

A third body of work addresses the generation not of interface widgets within a conversation, but of complete, deployable applications from natural language descriptions.
The term \emph{vibe coding}, coined by Karpathy~\cite{karpathy2025vibecoding} in early 2025, captures the emerging practice: a user describes what they want in natural language, accepts the AI-generated codebase without reviewing it in detail, and iterates by pasting errors back into the prompt.
A cohort of tools---Lovable~\cite{lovable2024}, Bolt~\cite{bolt2024}, and Vercel's v0~\cite{vercel2023v0}---has made this practice accessible to non-developers, generating full-stack applications, databases, and authentication systems from a single prompt.
The speed and accessibility of these tools has made them genuinely transformative for prototyping: Y Combinator managing partner Jared Friedman reported that 25\% of startups in its Winter 2025 batch had codebases that were 95\% AI-generated~\cite{mehta2025yc}.
What unifies this ecosystem is a shared conception of the artifact being produced: a software application is the \emph{goal} of the interaction, not a medium through which interaction continues.

Empirical work on how practitioners actually use these tools confirms this framing.
Chen et al.~\cite{chen2025genuistudy} conducted a week-long formative study with 37 UX-related professionals using a state-of-the-art GenUI tool, finding that participants treat the generated application as a deliverable---something to hand off to stakeholders or deploy for users---rather than as a substrate for ongoing collaborative work.
Subramonyam et al.~\cite{subramonyam2025vibecoding} reached a similar conclusion through interviews with 20 UX professionals: vibe coding tools are most readily adopted for early-stage prototyping and personal projects, with a clear transition point at which the generative conversation ends and the resulting application takes over.
In both studies, the conversational interface and the generated application occupy distinct, sequential roles: the conversation produces the app; the app is then used.
The two artifacts do not co-exist as parts of the same interaction.

This sequential structure reflects a deliberate architectural choice rather than an incidental limitation.
Jelly's authors note, in their critique of the code-generation approach to generative interfaces, that prompt-based revision of a codebase results in discontinuous transitions between generated states, with an opaque relationship between user intent and the resulting code that makes iterative control difficult~\cite{malleableui2025}.
The critique applies equally to vibe coding tools: because the generated application is conceived as a product, its internal structure is optimized for deployment rather than for communication.
The conversation that produced the app and the app itself have different audiences---developer and end-user, respectively---and this separation is by design.
SaC inverts this architecture.
The application is not the output of a prior conversation; it is the channel through which the current conversation proceeds.
Agent and human do not first converse and then use an app; they interact through the app, with each interaction cycle updating the application's state in a way that both parties can observe and act upon.
The contrast with vibe coding is therefore not one of capability---both paradigms rely on LLMs generating application code---but one of purpose: software as product versus software as medium.

A further distinction concerns accessibility and data grounding.
Vibe coding, despite its low technical barrier, still requires the user to arrive with \emph{product intent}---a prior conception of the artifact to be built---before the conversation can begin.
This is a non-trivial cognitive prerequisite that limits its reach to users who already think in terms of software artifacts.
SaC imposes no such prerequisite: agentic applications emerge as a byproduct of expressing intent naturally, without the user ever deciding to ``build something.''
Moreover, vibe coding produces empty containers that must be separately populated with data; in SaC, the application is natively grounded in the real task data that has already surfaced through dialogue, collapsing the gap between the tool and the information it is meant to organize.

%% file: sections/3.framework/3.1.The-Human-Agent-Environment-Interaction-Model.tex
\subsection{The Human--Agent--Environment Interaction Model}
\label{sec:interaction-model}


The emergence of capable AI agents introduces a fundamentally new structure to how humans interact with information.
We identify three distinct participants in this structure, each possessing complementary capabilities and limitations.

The \textbf{Human} ($H$) is the holder of intent---an evolving, often underspecified goal that motivates the interaction.
Humans possess domain judgment, contextual preferences, and the ability to make value-laden decisions that resist full automation.
However, human cognitive bandwidth is finite: we cannot simultaneously process large volumes of unstructured information, nor can we directly access, aggregate, or cross-reference data distributed across heterogeneous web sources.

The \textbf{Agent} ($A$) possesses knowledge and execution capabilities that complement human limitations: information retrieval, computation, multi-step reasoning, tool orchestration, and data synthesis.
Yet regardless of how capable agents become, two irreducible facts constrain fully autonomous action.
First, human preferences are not static inputs but \emph{evolve through the interaction itself}, and understandings are developed through exploration.
Second, in many consequential tasks, humans do not wish to delegate decisions entirely---not because the agent is untrustworthy, but because participation in the decision process carries intrinsic value: building understanding, maintaining agency, and preserving the capacity to change course.

The \textbf{Environment} ($E$) encompasses the external information sources and executable services that the agent can access on the human's behalf---web pages, databases, APIs, local file systems, sensor data, or any other digitally accessible resource or services.
This environment is vast but uncooperative: information is heterogeneous in format, distributed across organizational boundaries, dynamically changing, and never pre-organized for any particular user's particular task.

A critical observation follows from this characterization: \emph{no single participant can independently complete a non-trivial information task}.
The human has intent but lacks execution capability at scale; the agent has knowledge and execution capability but requires ongoing human input to track evolving preferences; the environment has information and service but lacks organization around any specific goal.
Productive outcomes emerge only through sustained, iterative collaboration among all three.


\paragraph{The interaction problem.}
Given three participants with complementary capabilities, the central design question becomes: \emph{what communicative structure best supports their collaboration?}

The dominant paradigm---the chat interface---implements this collaboration as a serial text channel between $H$ and $A$, with $A$ acting as a proxy for $H$'s access to $E$.
The resulting information flow takes the form:
\begin{equation}
H \xrightarrow{\text{NL}} A \xrightarrow{\text{Tools}} E \xrightarrow{\text{data}} A \xrightarrow{\text{NL}} H
\label{eq:chat-flow}
\end{equation}
This linear chain contains two natural language bottlenecks.
In the $H \rightarrow A$ direction, the user's parameterized intent---which may involve constraints on multiple dimensions, relative priorities, and conditional preferences---must be serialized into a natural language utterance.
In the $A \rightarrow H$ direction, structured execution results---tables, comparisons, spatial relationships, hierarchical data---must be flattened into sequential prose.


\paragraph{The cognitive case for structured interaction.}
The limitations of chat are not merely architectural---they are rooted in how human cognition processes information and formulates action.

The human visual system is inherently parallel: when presented with a spatially organized display, the eye extracts patterns, anomalies, and relationships through rapid scanning, often within hundreds of milliseconds.
Reading text, by contrast, is a fundamentally serial process: meaning must be decoded linearly, sentence by sentence, with the reader maintaining a mental model that is continuously updated but never directly visible.
This asymmetry goes beyond individual visual elements.
A graphical interface is not merely a collection of tables, cards, and charts---it is an \emph{information architecture}: a deliberate arrangement of hierarchy, grouping, spatial proximity, and visual weight that encodes relationships between pieces of information.
Which items are grouped together, what is foregrounded versus nested, how elements are spatially related---these architectural choices perform \emph{cognitive work} on behalf of the user by externalizing structure that would otherwise need to be mentally reconstructed from prose.
Linear text can describe such structure, but it cannot embody it; the reader must build the architecture internally, holding it in working memory while continuing to read.

Critically, structured interfaces do more than reduce the entropy of \emph{information presentation}---they simultaneously externalize the space of \emph{possible actions}.
A filter control communicates both a fact (``price is a relevant dimension'') and an operation (``you can constrain this dimension'') in a single visual element.
A sortable column header, a selectable item, a draggable boundary---each is simultaneously information and affordance, perception and action unified in one interface element.
In text, these two functions are necessarily separated: the information is in the prose, but the action must be independently conceived by the user and then articulated as a new message.
This separation forces a cognitive round-trip---read, interpret, formulate, type---that structured interfaces collapse into a single perceptual-motor act.


\paragraph{Deriving the mediating artifact.}
The preceding analysis---the structural constraints of chat identified in Section~\ref{sec:intro} and the cognitive advantages of spatial, action-integrated presentation---together define the requirements for a new mediating artifact.
We now ask: what properties would such an artifact need to resolve these constraints while leveraging the natural strengths of human perception?

\emph{Property 1: Structured Representation.}
To resolve representation mismatch, the agent needs the ability to present its output through spatial information architecture as a powerful complement to linear text.
This means not only choosing appropriate visual forms for individual data types---tables for comparison, maps for geospatial relations, timelines for temporal sequences---but composing them into a coherent layout where hierarchy, grouping, and spatial relationships encode the semantic structure of the task.
As established above, such architecture performs cognitive work that linear text cannot: it externalizes relationships and patterns so the user perceives rather than reconstructs them.

\emph{Property 2: Bounded Interaction.}
To resolve interaction entropy, the user's available actions at any given moment are concretized as discoverable interface affordances.
Rather than facing the unbounded action space of natural language (``what should I type next?''), the user encounters a contextually relevant set of operations---filters, selections, expansions, comparisons---that can be recognized and selected rather than recalled and articulated.
As noted above, each affordance unifies perception and action in a single element: the user simultaneously understands what is possible and can execute it, collapsing the cognitive round-trip of read--interpret--formulate--type into a direct perceptual-motor act.
This transforms the cognitive task of intent expression from recall to recognition, a distinction with well-established implications for interaction efficiency~\cite{nielsen1994usability}.

\emph{Property 3: Persistent, Bidirectional State.}
To resolve ephemeral interaction state, the artifact persists as a persistent structured object that both parties can continue to communicate upon directly, through the artifact itself, rather than through a separate text channel.
The agent shapes the artifact through rendering; the user shapes it through direct modification; and each party's actions are legible to the other through the artifact's evolving state.
The artifact is not a display surface adjacent to a conversation, but the conversation medium itself.

An artifact satisfying all three properties is neither a pure conversational message, nor a static document, nor a pre-built application.
It is supposed to be a dynamically generated, interactive application---constructed by the agent in response to the user's intent, presenting structured information through spatial arrangement, exposing contextually relevant actions as interface affordances, and persisting as a shared, manipulable workspace across interaction cycles.
We term such artifacts \textbf{agentic applications}, and the paradigm in which they serve as the primary medium of human--agent interaction \textbf{Software as Content} (SaC).

\begin{figure*}[t]
  \centering
  \includegraphics[width=\textwidth]{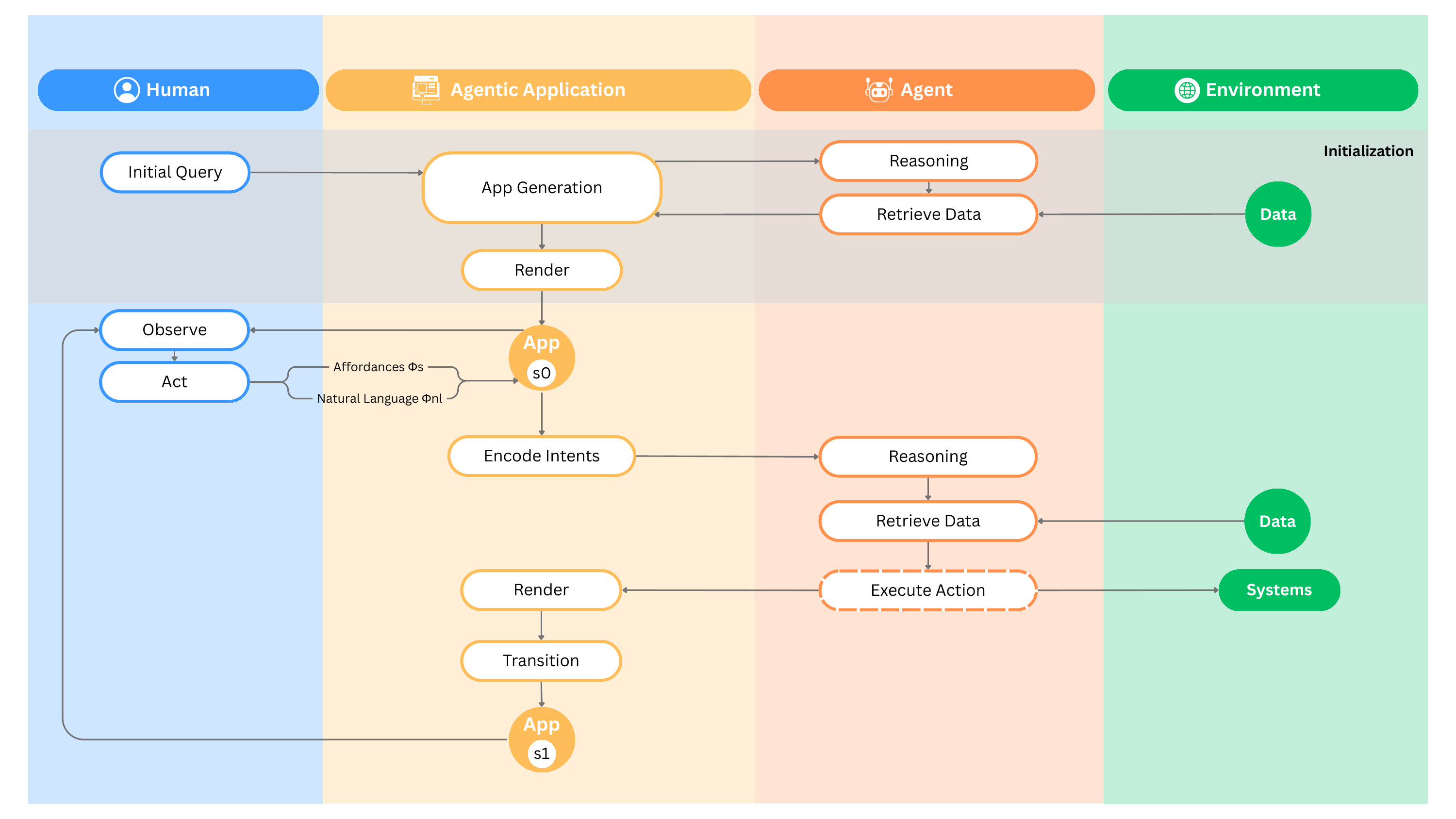}
  \caption{The SaC interaction cycle across four participants.
  \textit{Initialization} (top): the user's initial query triggers App Generation,
  in which the agent reasons over the request, retrieves data from the environment,
  and constructs the initial application state~$s_0$.
  \textit{Interaction cycle} (center): the human observes~$s_0$ and acts through
  one of two input channels---structured affordances~$\Phi^s$ or the natural
  language channel~$\Phi^{nl}$---whose intents are encoded and dispatched to the
  agent. The agent reasons, optionally retrieves data or executes actions against
  external systems, and renders an update~$\Delta$ that drives the transition
  $s_{t+1} = s_t \oplus \Delta$. The resulting state~$s_1$ is presented back to
  the human, closing the loop. Dashed borders denote optional operations.
  When the agent determines that a plain text response suffices
  (e.g., for simple factual queries or brief clarifications), the render step
  produces a text reply directly without instantiating or updating an agentic
  application; the cycle does not proceed.}
  \label{fig:hae}
\end{figure*}

%% file: sections/3.framework/3.2.Agentic-Applications-as-a-Content-Modality.tex
\subsection{Agentic Applications as a Content Modality}
\label{sec:agentic-app}


The preceding section derived three properties that a mediating artifact for human--agent--environment collaboration would need to satisfy, and identified agentic applications as the class of artifacts that fulfills them.
We now formalize what agentic applications are and how they operate, then situate them within the broader landscape of content and interaction modalities.


\paragraph{The interaction cycle.}
At any point in an interaction, an agentic application exists in a state $s_t$ comprising three components:
\begin{equation}
s_t = (V_t,\; \Phi_t,\; C_t)
\label{eq:state}
\end{equation}
where $V_t$ denotes the \emph{view}---the agent's rendered output, which may range from plain text for simple responses to full information architecture with spatial layout, visual hierarchy, and interactive elements, depending on the complexity and structure of the task;
$\Phi_t$ denotes the \emph{affordance set}, which comprises two complementary components:
\begin{equation}
\Phi_t = \Phi_t^s \cup \Phi_t^{nl}
\label{eq:affordance}
\end{equation}
Here $\Phi_t^s$ is the set of \emph{structured affordances}---contextually generated interface operations such as filters, selectors, sort controls, and action triggers---that the agent produces based on its model of the current task state.
$\Phi_t^{nl}$ is the \emph{natural language channel}---an always-available input through which the user can express any intent that $\Phi_t^s$ does not cover.
$\Phi_t^{nl}$ includes both free-form text input and \emph{anticipatory affordances}: generated follow-up intents that surface predicted next steps as recognizable options, reducing the cognitive cost of natural language expression from recall to recognition.
$\Phi_t^s$ is finite and dynamically generated; $\Phi_t^{nl}$ is unbounded and permanently present.
Finally, $C_t$ denotes the \emph{agent context}---the execution state maintained by the agent, including retrieved data, inferred user preferences, task progress, and interaction history.

A single interaction cycle proceeds as follows:
\begin{enumerate}
    \item \textbf{Observe.} Human perceives the current view $V_t$, discovers available actions through $\Phi_t^s$ and anticipatory affordances within $\Phi_t^{nl}$, while also retaining the option to express arbitrary intent through free-form input.
    \item \textbf{Act.} The human either selects a structured operation $a \in \Phi_t^s$, or issues a natural language input through $\Phi_t^{nl}$.
    \item \textbf{Encode.} The action is translated into an event $e$. A structured operation yields a parameterized event directly (e.g., \texttt{\{action: ``filter'', field: ``price'', operator: ``<'', value: 500\}}); a natural language input yields a free-form intent signal that the agent interprets in context.
    \item \textbf{Execute.} The agent receives $e$, interprets it within the current context $C_t$, and determines an execution plan---which may involve querying $E$, performing computation, or reorganizing existing data.
    \item \textbf{Retrieve.} The agent interacts with the environment $E$ as needed, obtaining new data $d$ or confirming existing information.
    \item \textbf{Render.} The agent constructs a state update $\Delta = \textsc{render}(V_t, C_t, e, d)$, determining how the view, affordance set, and context should change in response to the user's action and the newly obtained information.
    \item \textbf{Transition.} The application transitions to a new state:
    \begin{equation}
    s_{t+1} = s_t \oplus \Delta = (V_{t+1},\; \Phi_{t+1},\; C_{t+1})
    \label{eq:transition}
    \end{equation}
\end{enumerate}

Several aspects of this cycle merit emphasis.

First, $\Phi_t$ is a hybrid of bounded and unbounded interaction.
The structured affordances $\Phi_t^s$ are the agent's best prediction of what the user might want to do next---a finite, contextually relevant set of low-cognitive-cost operations.
But $\Phi_t^s$ is necessarily an approximation: the agent cannot anticipate every direction the user's intent may evolve.
The natural language channel $\Phi_t^{nl}$ completes the picture, ensuring the user is never constrained to only what the agent has foreseen.
The key design goal of a SaC system is to minimize the user's dependence on unaided recall: maximizing the coverage of $\Phi_t^s$ so that the most frequent operations are directly parameterized, and supplementing $\Phi_t^{nl}$ with anticipatory affordances so that even when intent must flow through the natural language path, the user can recognize and select rather than independently formulate.

Second, the two input paths produce events at different levels of structure.
When a user acts through $\Phi_t^s$---adjusting a filter, selecting items for comparison, toggling a view---the resulting event is inherently parameterized, carrying precise semantic content without requiring the user to articulate it verbally.
When a user acts through $\Phi_t^{nl}$---whether by typing freely or by selecting an anticipatory affordance---the agent receives a free-form intent signal that it interprets in the context of $C_t$ and $V_t$.
Both paths resolve the $H \rightarrow A$ bottleneck in Equation~\ref{eq:chat-flow}, but $\Phi_t^s$ does so with lower cognitive cost and higher semantic precision ~\cite{horvitz1999principles}.

Third, $\textsc{render}(\cdot)$ is both dual-channel and state-aware.
For simple queries---factual lookups, greetings, brief clarifications---the agent may respond with plain text, just as in a conventional chat.
For tasks involving structured data, multi-dimensional comparison, or iterative exploration, the agent constructs or updates an information architecture.
Note that $\textsc{render}$ takes the current view $V_t$ as an explicit input: rather than regenerating the interface from scratch, the agent selects an update strategy based on the structural compatibility between the new information and the existing interface.
If the new data is structurally compatible with the current layout (e.g., additional items of the same type), elements are added without altering the existing architecture.
If the data introduces a related but structurally distinct facet (e.g., a sub-topic requiring a different organization), a new view is appended within the existing application (e.g., as an additional tab or panel).
If the new direction is sufficiently divergent, a new agentic application may be initiated entirely.
This resolves the $A \rightarrow H$ bottleneck in Equation~\ref{eq:chat-flow} not by eliminating text, but by making it one option within a richer, state-aware output repertoire.

Fourth, the artifact is a \emph{persistent, bidirectional medium}.
The agent shapes $s_t$ through rendering; the user shapes $s_{t+1}$ through both structured manipulation and natural language input.
Both parties operate on the same shared object, and the artifact's evolving state serves as a legible record of their collaboration---each cycle building upon the last rather than starting from a blank message.

Fifth, the two modalities---structured interaction and natural language---are \emph{permanently complementary}.
This distinguishes SaC from the historical GUI--CLI transition, where both modalities were predefined and finite, and GUI largely subsumed CLI by offering a more discoverable interface to the same fixed capabilities.
In the SaC case, both modalities are generative and unbounded---neither subsumes the other.
SaC expands the bandwidth of human--agent collaboration rather than replacing one medium with another.


\paragraph{Positioning in the modality landscape.}
Agentic applications are not the first attempt to move beyond linear text in human--agent interaction.
To understand what is genuinely new about the SaC paradigm, it is useful to situate agentic applications within the broader landscape of interaction modalities---distinguishing them not only from chat, but from traditional software and current generative UI systems.

Table~\ref{tab:modality} compares four modalities across four dimensions derived from the interaction model above.
The \emph{Generation} dimension distinguishes dynamically generated artifacts from pre-built ones.
The \emph{H$\rightarrow$A Input} dimension characterizes how the human directs the system: through interactive affordances, through natural language, or both.
The \emph{A$\rightarrow$H Output} dimension characterizes how the system presents information: as spatial information architecture or as linear text.
The \emph{Evolution} dimension captures whether the artifact progressively specializes to the user's task over successive cycles.

\begin{table}[h]
\centering
\caption{Positioning of agentic applications within the interaction modality landscape.}
\label{tab:modality}
\begin{tabular}{lcccc}
\toprule
& \textbf{Generation} & \textbf{H$\rightarrow$A Input} & \textbf{A$\rightarrow$H Output} & \textbf{Evolution} \\
\midrule
Traditional software        & Pre-built  & Interactive           & Spatial   & None         \\
Chat                        & Dynamic    & NL                    & Linear    & None         \\
Current GenUI               & Dynamic    & NL                    & Spatial   & None         \\
\textbf{Agentic apps (SaC)} & Dynamic    & Interactive + NL      & Spatial   & Progressive  \\
\bottomrule
\end{tabular}
\end{table}

A key structural property not captured by this table deserves emphasis: in agentic applications, the artifact itself is the interaction medium.
User actions---filtering, selecting, rearranging---flow through the application directly to the agent as structured events, and agent responses flow back as updates to the same shared object.
This is fundamentally different from current GenUI systems, where the artifact persists but remains a \emph{view-only} surface: the user must leave the artifact, return to a separate text input, and describe modifications in natural language.
The artifact in SaC is not adjacent to the conversation---it is the conversation.

Two further observations follow from the table.
First, traditional software and agentic applications share the interactive affordance model and spatial output---but traditional software is pre-built, encoding a fixed developer's model of what users might need.
Agentic applications generate both the interface and the affordances dynamically, in response to this user's intent and the data retrieved from the environment.
The interaction model is the same; the generative mechanism is fundamentally different.

Second, current GenUI systems resolve the A$\rightarrow$H output bottleneck by producing spatial information architecture rather than linear text.
But they leave the H$\rightarrow$A direction unchanged: the user's next action is still a natural language message typed into an unconstrained text box.
The distinction between SaC and current GenUI is therefore not one of degree---it is structural.
Current GenUI improves the agent's output format; SaC changes the interaction architecture.

%% file: sections/3.framework/3.3.Lifecycle.tex
\subsection{The SaC Lifecycle}
\label{sec:lifecycle}

\begin{figure}[tp]
  \centering
  \includegraphics[height=0.9\textheight, keepaspectratio]{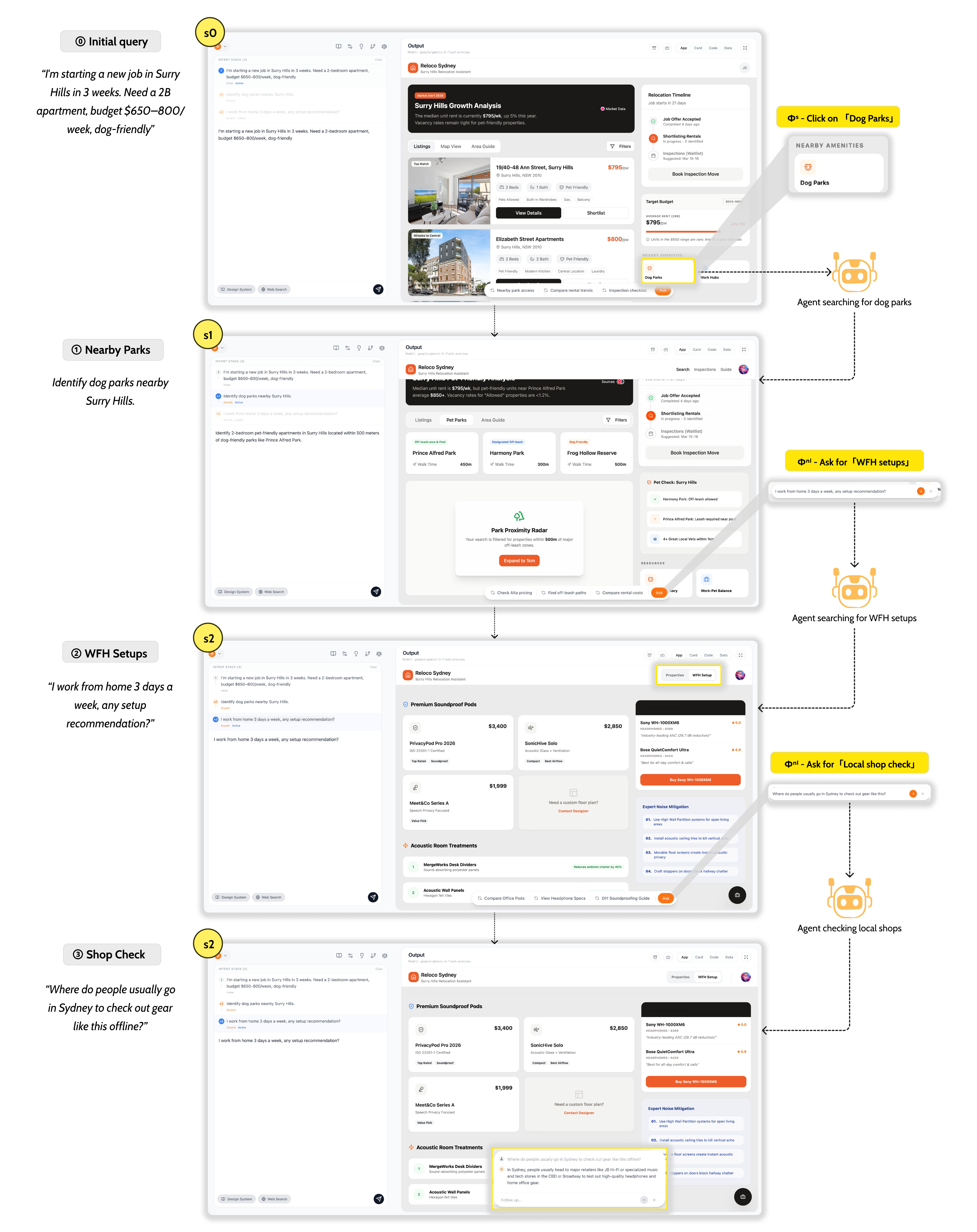}
  \caption{The primary SaC application evolution illustrated through a Sydney relocation
  scenario (\textit{Reloco Sydney}).
  Each panel shows the interaction trace (left, intent stack) alongside the current
  agentic application state (right).
  \textit{Cold start}~($s_0$): the user's initial query is dispatched to the agent,
  which retrieves market and listing data and constructs an initial application
  comprising a property listing view, a relocation timeline, and contextually
  relevant affordances.
  \textit{First cycle}~($s_0 \rightarrow s_1$): the user clicks the ``Dog Parks''
  element~($\Phi^s$), dispatching an intent to the agent; the agent retrieves
  proximity data and updates the application by adding a Pet Parks tab and a
  Park Proximity Radar panel.
  \textit{Second cycle}~($s_1 \rightarrow s_2$): the user submits a natural language
  query~($\Phi^{nl}$)---``I work from home 3 days a week, any setup
  recommendation?''---dispatching a new intent; the agent retrieves relevant
  information and extends the application with a WFH-focused acoustic setup section.
  \textit{Plain text reply}: a subsequent factual query about local stores is answered
  directly by the agent without updating~$s_2$, illustrating that the SaC cycle does
  not proceed when a text response suffices.}
  \label{fig:lifecycle}
\end{figure}

While a single interaction cycle is the atomic unit of the model, the defining characteristic of SaC lies in the cumulative effect of successive cycles.
Each state transition $s_t \rightarrow s_{t+1}$ carries information from all preceding interactions.
The view $V_{t+1}$ reflects not just the latest query result, but the accumulated refinements of $t$ previous cycles.
The affordance set $\Phi_{t+1}^s$ adapts to the trajectory of the interaction, surfacing operations that are relevant given what the user has already explored and decided.
The context $C_{t+1}$ incorporates the agent's growing model of this user's preferences and priorities for this task.

Over the course of $n$ cycles, the application undergoes \emph{progressive specialization}: it evolves from a generic, intent-matched starting point toward a configuration that is specifically tailored to this user's needs for this task.
The initial state $s_0$, generated from the user's first expression of intent, is necessarily general---the agent must make assumptions in the absence of detailed preference information.
But each subsequent cycle narrows the space: irrelevant options are filtered out, preferred dimensions are foregrounded, useful intermediate results are preserved, and the information architecture increasingly reflects the user's mental model of the task.

In the limit, a sufficiently evolved agentic application converges toward what we term \emph{personalized software}---a task-specific tool that, had it been designed by a human developer with full knowledge of this user's requirements, might have been built as a dedicated application.
The crucial difference is that this personalized tool was not pre-designed; it emerged from the interaction itself.
This distinguishes SaC from AI-assisted software construction approaches such as vibe coding (\S\ref{sec:related:softwaregen}), where the generated application is the end product; in SaC, the application is an interaction medium whose personalization is a byproduct of use, not the goal of construction.

\paragraph{Initial generation.}
The lifecycle begins with a cold start: the user's first expression of intent, and the agent's construction of $s_0$.
This is a qualitatively different moment from subsequent cycles, because there is no prior state to update---the agent must generate an entire application from a single, often underspecified input.
The core decision at this stage is whether an agentic application is warranted at all: a simple factual query or a brief clarification may be better served by a plain text response.
When the task does call for structured interaction, the agent selects an initial information architecture and populates a starting affordance set $\Phi_0^s$ based on the inferred intent and the data available from the environment.
The types of information architecture available to the agent are characterized in \S\ref{sec:taxonomy}.

\paragraph{Evolution trajectories.}
Progressive specialization describes the general trend, but the actual trajectory of an agentic application is rarely a smooth linear progression.
In practice, the evolution exhibits recurring patterns of \emph{convergence} and \emph{divergence}.

Convergence occurs when the user's actions follow the trajectory anticipated by the current view---selecting from exposed affordances, refining within the existing information architecture, linearly advancing toward the task goal.
Divergence occurs when the user's intent moves beyond what the current view covers---triggered either by agent-suggested exploratory follow-ups or by the user's own natural language input expressing an unanticipated direction.
It is the interplay of these two modes that drives the application's structural growth over time.
Figure~\ref{fig:lifecycle} illustrates this interplay through a concrete relocation assistance scenario.
The three render update strategies defined in \S\ref{sec:agentic-app} map directly onto these trajectory patterns.
Convergence manifests as element-level updates within a stable layout---the information architecture stays fixed while the data within it is refined.
Divergence manifests as structural growth---new tabs, panels, or views are appended to accommodate the newly opened direction.
Radical divergence---where the user's intent shifts to an entirely different task---may trigger the creation of a new agentic application altogether.

The lifecycle of an agentic application is therefore not a single arc from general to specific, but a tree-like structure: a trunk of progressive specialization with branches where the user explored alternative directions, some of which were pruned and others retained.

\paragraph{Retirement and reuse.}
An agentic application reaches the end of its active lifecycle through several possible paths.
The most common is simply task completion: the user has obtained the information they needed or executed the desired action, and the application is abandoned---much as a chat conversation is closed once the user's question has been answered.
As a content modality, this disposability is a feature, not a limitation---just as a user closes a web page after reading it, an agentic application can be discarded after it has served its purpose.

However, the SaC paradigm also supports longer-lived outcomes.
A sufficiently specialized application---one that has evolved into a well-tailored tool through many cycles of interaction---may be worth preserving.
The user might save it as a reusable template: a personalized dashboard, a curated comparison workspace, or a configured monitoring view that can be revisited later.
Such applications can also be shared: a user who has refined an agentic application for a particular task can distribute it to others, who receive not just information but a ready-made interactive tool that they can continue to iterate and evolve for their own needs.
This opens a natural extension of the SaC paradigm toward agentic application marketplaces and content platforms---a direction we discuss further in \S\ref{sec:futurework}.

In this sense, SaC closes a loop that traditional software leaves open: where conventional applications are designed generically and then customized through settings, agentic applications are generated specifically and can be promoted to persistent, shareable tools if they prove valuable enough to retain.

%% file: sections/3.framework/3.4.A-Taxonomy-of-Agentic-Applications.tex
\subsection{A Taxonomy of Agentic Applications}
\label{sec:taxonomy}

The interaction cycle defined in \S\ref{sec:agentic-app} is general---it describes how any agentic application operates.
But agentic applications are not monolithic: a flight comparison tool, a research dashboard, and a step-by-step visa application guide differ substantially in their information architecture, affordance design, and evolution patterns.
This section characterizes the dimensions along which agentic applications vary, providing a framework for understanding and designing different application types.
We focus specifically on cases where the agent has determined that structured interaction is warranted; simple text responses, as discussed in \S\ref{sec:agentic-app}, fall outside this taxonomy.

\paragraph{Three dimensions of variation.}
The form of an agentic application is not determined by any single factor.
Drawing from the interaction model established in \S\ref{sec:interaction-model}, we observe that agentic applications vary along three largely independent dimensions: the \emph{structure of the user's intent}, the \emph{structure of the data} retrieved from the environment, and the \emph{temporal scope} of the interaction.

The first two dimensions jointly determine the application's form at any given moment.
The intent axis operates top-down: given what the user wants to accomplish, what kind of interaction does the application need to support?
The data axis operates bottom-up: given the information actually available, what form should the presentation and affordances take?
The two axes meet in the agent's decision space---intent determines the functional constraints (what the application must do), while data determines the concrete realization (how it looks and behaves).
The third dimension---temporal scope---determines the lifecycle shape: how long the application persists and how deeply it specializes.

\paragraph{Intent structure.}
We identify five broad categories of user intent, each implying a distinct pattern of information architecture and affordance design:

\emph{Selection}---choosing among multiple options.
The user needs to compare, filter, and evaluate a set of candidates (e.g., flights, restaurants, apartments).
The information architecture emphasizes parallel presentation of comparable items, with affordances for filtering, sorting, and side-by-side comparison.
The dominant information flow is A$\rightarrow$H: the agent surfaces options, the user narrows.

\emph{Exploration}---understanding a complex information space.
The user seeks to build a mental model of a topic, often without a precise goal at the outset ~\cite{marchionini2006exploratory} (e.g., researching a medical condition, exploring investment options, surveying a new field).
The information architecture emphasizes hierarchical organization and progressive disclosure, with affordances for drilling down, expanding related topics, and bookmarking.
Information flow is bidirectional throughout: the user's exploration signals continuously reshape what the agent surfaces next.

\emph{Execution}---completing a multi-step workflow.
The user has a defined goal that requires a sequence of actions (e.g., filing an application, configuring a service, planning a trip).
The information architecture emphasizes sequential progression with clear step indicators, with affordances for form input, validation, and navigation between steps.
The dominant information flow is H$\rightarrow$A: the user provides input, the agent processes and advances.

\emph{Monitoring}---tracking state changes over time.
The user needs ongoing visibility into evolving data (e.g., portfolio performance, project status, competitor activity).
The information architecture emphasizes dashboards with key metrics, alerts, and trend visualizations, with affordances for adjusting thresholds, time ranges, and notification preferences.
The application typically persists across sessions, linking this intent type to the longer-lived end of the lifecycle spectrum described in \S\ref{sec:lifecycle}.

\emph{Creation}---producing or editing content.
The user is generating an artifact with the agent's assistance (e.g., writing a document, designing a layout, composing code).
The information architecture centers on an editable workspace with real-time preview, with affordances for direct manipulation, version control, and iterative refinement.
Information flow is tightly coupled: each user edit triggers agent response, and each agent suggestion invites user modification.

\begin{table}[h]
\centering
\caption{Five intent categories and their characteristic interaction properties.}
\label{tab:intent-taxonomy}
\begin{tabular}{lllc}
\toprule
\textbf{Intent} & \textbf{Information Architecture} & \textbf{Primary Affordances} & \textbf{Information Flow} \\
\midrule
Selection    & Parallel items          & Filter, sort, compare          & $A \rightarrow H$ dominant \\
Exploration  & Hierarchical, progressive & Drill-down, expand, bookmark  & Bidirectional \\
Execution    & Sequential steps        & Form input, validation         & $H \rightarrow A$ dominant \\
Monitoring   & Dashboard, metrics      & Thresholds, time range, alerts & $A \rightarrow H$ continuous \\
Creation     & Editable workspace      & Direct manipulation, preview   & Tightly coupled \\
\bottomrule
\end{tabular}
\end{table}

These categories are not mutually exclusive---a single agentic application may transition between them as the task evolves, or combine them simultaneously (e.g., a travel planning app that involves selection of flights, exploration of destinations, and execution of bookings).
The lifecycle trajectories described in \S\ref{sec:lifecycle} often correspond to shifts between intent categories.

\paragraph{Data structure.}
The intent axis determines the functional pattern; the data axis determines its material realization.
The same intent category can produce very different interfaces depending on the structure of the information retrieved from the environment.

Consider selection intent applied to three domains: comparing flights produces a structured table (many items, uniform fields, numerical attributes), comparing restaurants produces an image-rich card grid (fewer items, visual-heavy, heterogeneous attributes), and comparing insurance plans produces a feature-matrix with expandable detail panels (dense text, conditional clauses, hierarchical sub-features).
All three share the same functional constraints---parallel presentation, dimension control, decision affordances---but the components, layout, and interaction details differ substantially.

This observation has a direct design implication: the mapping from intent to interface cannot be a fixed recipe.
A selection pattern is not ``a table with filters''; it is a set of functional constraints (items are comparable, dimensions are switchable, a decision exit exists) that the agent instantiates differently depending on what the data actually looks like.
The agent's task is to match the functional constraints of the intent pattern to the structural properties of the available data---a process we detail in the system architecture (\S\ref{sec:system}).

\paragraph{Interaction time span.}
The third dimension---orthogonal to both intent and data---captures the expected temporal scope of the interaction.
At one extreme, some generated interfaces are single-turn: a currency comparison chart, a weather forecast with hourly breakdown, a nutritional summary.
These involve structured presentation but no subsequent interaction cycle---they are, in effect, generative UI components, and represent the boundary where SaC reduces to conventional GenUI.
Session-length applications---a research exploration, a comparison workflow, a trip planning process---undergo progressive specialization and exhibit the convergence--divergence patterns described in \S\ref{sec:lifecycle}.
Long-lived applications---monitoring dashboards, portfolio trackers, competitive intelligence views---represent the fully specialized endpoint: instances that have been promoted from disposable content to persistent, personalized tools.
The three points on this temporal spectrum---single-turn, session-length, and long-lived---correspond to fundamentally different lifecycle shapes and different degrees of progressive specialization.

%% file: sections/3.framework/3.5.Design-Principles-for-SaC-Systems.tex
\subsection{Design Principles for SaC Systems}
\label{sec:principles}

The interaction model, lifecycle, and taxonomy developed in the preceding sections are descriptive---they characterize what agentic applications are and how they behave.
This section derives a set of prescriptive principles that follow from the model and guide the design of SaC systems.
Each principle is traced to a specific element of the formal framework; together, they define the design space within which concrete implementations operate.

\paragraph{Principle 1: Structured interaction first, natural language always.}
The dual-channel affordance model ($\Phi_t = \Phi_t^s \cup \Phi_t^{nl}$) implies a clear design priority: maximize the coverage of $\Phi_t^s$ so that the most common and most parameterizable user actions can be expressed through structured affordances, while ensuring $\Phi_t^{nl}$ remains permanently available for everything else.
The goal is not to eliminate natural language but to reduce the frequency with which users are forced to rely on it---to make the structured path the path of least resistance for routine operations, while preserving the expressive power of free-form input for novel, complex, or unanticipated needs.

\paragraph{Principle 2: Match output modality to task structure.}
The view $V_t$ ranges from plain text to full information architecture, and the agent's choice along this spectrum is itself a design decision.
Not every agent response warrants a generated application: a factual answer, a brief clarification, or a simple acknowledgment is better served by text.
Generating an interactive interface where a sentence would suffice imposes unnecessary cognitive overhead; conversely, flattening structured information into prose wastes the spatial and interactive capacity of the medium.
The agent should select the lightest modality that faithfully represents the structure of the response.

\paragraph{Principle 3: Evolve, don't regenerate.}
The render function takes the current view $V_t$ as an explicit input, and the transition operator ($s_{t+1} = s_t \oplus \Delta$) is defined as an incremental update rather than a wholesale replacement.
This is not merely an implementation choice---it reflects a cognitive reality: the user has built a mental model around the current interface, and regenerating it from scratch destroys that accumulated context.
Incremental evolution preserves the user's orientation while integrating new information, following the three-level strategy described in \S\ref{sec:agentic-app}: update elements within a stable layout when the new data is structurally compatible, extend the application with new views when the data introduces a new facet, and initiate a new application only when the direction is fundamentally divergent.

\paragraph{Principle 4: Constrain semantics, not components.}
The taxonomy in \S\ref{sec:taxonomy} establishes that the same intent pattern produces different interfaces depending on the data structure.
This has a direct implication for how design knowledge is encoded in SaC systems: constraints should be specified at the level of \emph{semantic goals} (what the interface must accomplish---parallel comparison, dimension switching, decision exit) rather than \emph{component prescriptions} (use a table, place filters on the left, render cards in a grid).
Semantic-level constraints complement the generative capabilities of the agent, which brings its own substantial knowledge of effective UI patterns from training data; component-level prescriptions, while sometimes providing useful guidance, inevitably limit the diversity of generated interfaces and can conflict with the agent's own judgment about what best fits the data.
This does not mean component-level design systems have no role---standardized component libraries (providing consistent styling, accessibility, and interaction primitives) serve as valuable building blocks.
But the selection and composition of those components should be guided by semantic constraints and data structure, not dictated by rigid templates.
The design system for a SaC implementation therefore defines \emph{patterns as functional constraint sets}, not as fixed recipes, leaving the agent to select the concrete realization that best fits the data at hand.

\paragraph{Principle 5: Preserve human agency.}
The interaction model in \S\ref{sec:interaction-model} identifies two irreducible facts about the human participant: preferences evolve through interaction, and humans in many contexts do not wish to fully delegate decisions.
These facts impose a design obligation: the interface should keep the user informed, in control, and free to change course at any point.
Concretely, this means the information presented should be traceable to its sources (the user can verify where the data came from), the agent's choices should be overridable (the user can modify the information architecture, not just the data within it), and the natural language channel $\Phi_t^{nl}$ should always be available as an exit from any structured interaction path.
As agent capabilities continue to advance, the temptation to automate more of the decision process will grow; this principle serves as a reminder that the value of SaC lies in augmenting human judgment, not replacing it.

\paragraph{Principle 6: Manage accumulated complexity.}
The transition operator $s_{t+1} = s_t \oplus \Delta$ is additive by nature: each cycle contributes new information, new views, and new affordances to the application state.
Over extended interactions, this accumulation poses a structural challenge that chat-based systems do not face---the very persistence and structure that give SaC its cognitive advantages can, if left unmanaged, produce an interface whose complexity overwhelms those advantages.
SaC systems should therefore actively manage state growth at three levels: at the interface level, by folding inactive branches, archiving completed sub-tasks, and foregrounding currently relevant views; at the context level, by compressing interaction history into summary representations that preserve key decisions and preference signals while discarding intermediate steps; and at the application level, by treating the fork-to-new-app strategy described in \S\ref{sec:agentic-app} as a natural complexity boundary---a point at which it is better to start fresh than to continue extending an overloaded structure.
The specific thresholds and strategies for complexity management remain an open design question; we discuss this further in \S\ref{sec:limitations}.

%% file: sections/4.system-architecture/4.1.Overview.tex

\subsection{Overview}
\label{sec:system:overview}

\begin{figure*}[t]
  \centering
  \includegraphics[width=\textwidth]{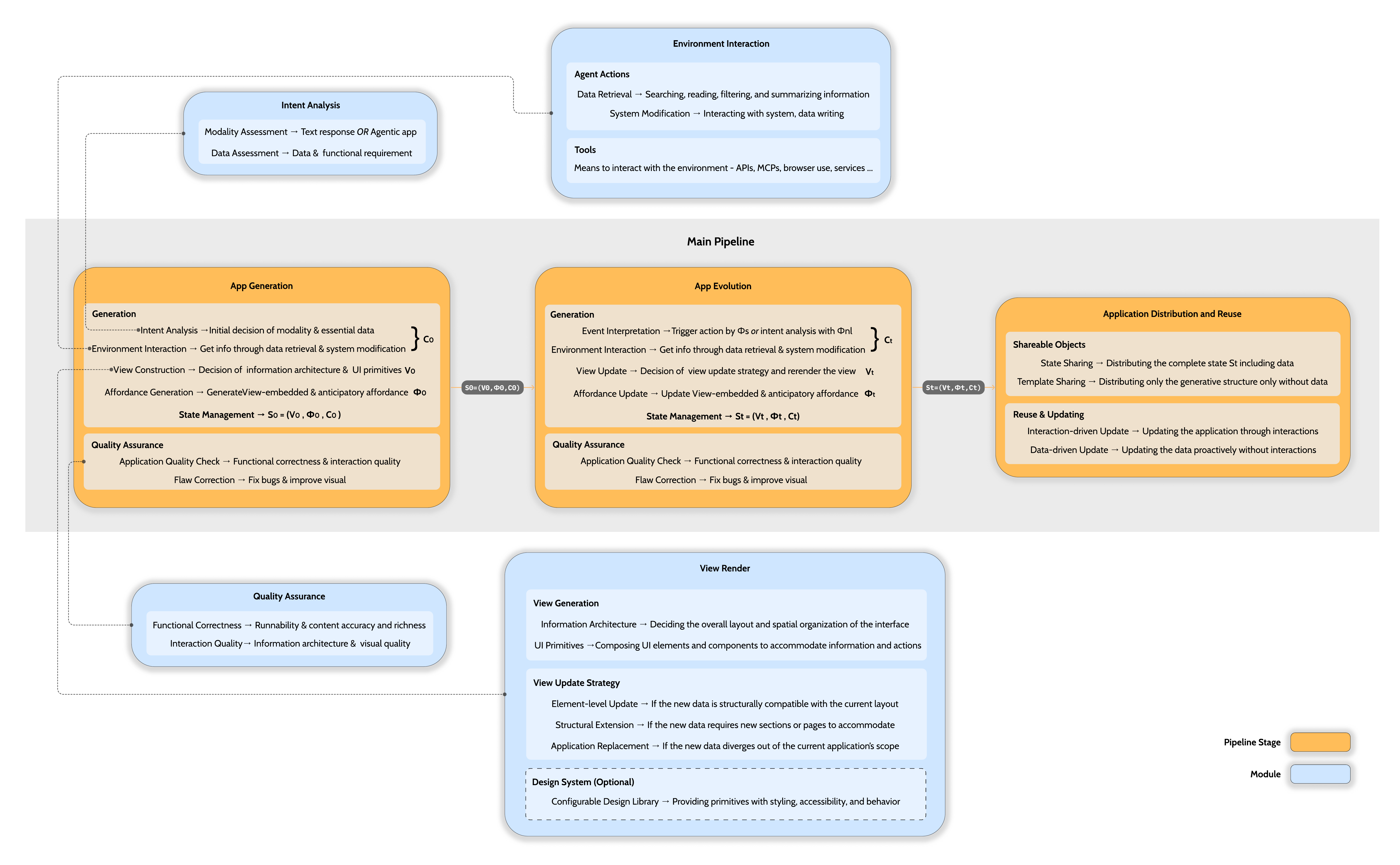}
  \caption{System architecture overview. The main pipeline (center) comprises three sequential stages: 
  \textit{App Generation} constructs the initial application state $s_0 = (V_0, \Phi_0, C_0)$ from a cold start; 
  \textit{App Evolution} computes successive state transitions $s_{t+1} = s_t \oplus \Delta$ in response to incoming events; 
  \textit{Application Distribution and Reuse} handles sharing and refresh of completed application states. 
  Four shared modules (outer boxes) are invoked across stages: 
  \textit{Intent Analysis} provides modality assessment and data requirements at every stage; 
  \textit{Environment Interaction} handles data retrieval and system modification on behalf of both generation and evolution; 
  \textit{View Render} is the shared computational primitive underlying both stages, 
  instantiating $V_0$ during generation and computing $\Delta$ during evolution, 
  with view update strategy selection (element-level update, structural extension, or application replacement) 
  determining the degree of structural change at each cycle;
  \textit{Quality Assurance} enforces functional correctness and interaction quality within each pipeline stage.}
  \label{fig:pipeline}
\end{figure*}

The system architecture distinguishes two kinds of components: a sequential pipeline that moves application state forward, and a cross-cutting capability that all pipeline stages depend on.

\paragraph{Cross-cutting: intent analysis.}
Intent analysis is not a single pipeline stage but a shared interpretive function invoked throughout the system.
At every point where the system must act---constructing $s_0$, computing a render update $\Delta$, or evaluating output quality---it first interprets the current intent: inferring the applicable intent category from \cref{sec:taxonomy}, making a preliminary modality assessment and identifying the data requirements and information architecture appropriate to the moment.
Because user intent evolves across interaction cycles rather than being fixed at the first utterance, this interpretive step cannot be performed once and cached; it recurs at each stage with access to the current context $C_t$.

\paragraph{Sequential pipeline.}
\Cref{fig:pipeline} illustrates the pipeline that carries application state from initial generation through to distribution.
Given a user's first expression of intent, the application generation stage (\cref{sec:system:generation}) constructs $s_0 = (V_0, \Phi_0^s, C_0)$ from scratch---building the initial view, populating the structured affordance set, and initializing the agent context.
Subsequent interaction events, whether originating from $\Phi_t^s$ or the natural language channel $\Phi_t^{nl}$, enter the application evolution stage (\cref{sec:system:evolution}), which implements the render function and produces $s_{t+1} = s_t \oplus \Delta$.
Both stages are gated by quality assurance (\cref{sec:system:qa}), which enforces a minimum quality floor before any application state is presented to the user.
The distribution and consumption layer (\cref{sec:system:distribution}) then handles how completed agentic applications are surfaced and made available for reuse.

\paragraph{Unifying abstraction.}
The render function is the central computational primitive of the pipeline: every user-visible state transition passes through it, including the construction of $s_0$, which is handled by invoking render against an empty prior state.
The generation and evolution stages therefore share a single underlying abstraction, differing only in whether a prior state exists.

%% file: sections/4.system-architecture/4.2.App-Generation.tex

\subsection{App Generation}
\label{sec:system:generation}

The construction of $s_0$ is qualitatively different from all subsequent state transitions.
In every later cycle, the render function has a prior state $s_t$ to build upon---existing structure to preserve, affordances to update, a context already seeded with retrieved data and inferred preferences.
At $s_0$, none of this exists: the system receives a single, often underspecified utterance and must construct a complete application state from scratch.
This cold-start problem unfolds across four steps: intent analysis, environment interaction, application construction, and affordance generation.

\paragraph{Intent analysis.}
Intent analysis first makes a preliminary modality assessment: given the intent type, is a structured agentic application likely to be warranted, or is a plain text response more appropriate (\cref{sec:principles})?
For tasks that the preliminary assessment suggests warrant structured interaction, intent analysis then produces a generation plan with two components.
The first is an intent category, drawn from the taxonomy in \cref{sec:taxonomy}, which determines the functional constraints the application must satisfy---what kinds of comparison, exploration, or execution the interface needs to support.
The second is a data assessment: what information must be retrieved from the environment, and which queries or API calls are needed to obtain it.
Critically, the data assessment does not determine what the interface will look like---the structure, density, and heterogeneity of the retrieved data are not known until retrieval actually completes, and are often unpredictable from the intent alone.
The intent category establishes functional constraints on the application, but the concrete information architecture of $V_0$ is decided only after the real data is in hand.

\paragraph{Environment interaction.}
Before construction can begin, the agent interacts with the environment $E$ to obtain the content that will ground the response.
This interaction is not limited to read operations: while selection and exploration intents primarily involve retrieving and filtering information, execution intents require the agent to act on or modify $E$---submitting forms, writing data, or triggering external services that change state.
In both cases, the interaction produces a return value: retrieved data for read operations, and an execution result or status confirmation for write operations.
The structure and complexity of this return value finalizes the modality decision: a simple status confirmation may warrant nothing more than a plain text response, while a rich, heterogeneous result set calls for a structured information architecture.

\paragraph{View construction.}
Given the generation plan and the return value from environment interaction, the agent constructs $V_0$.
This involves two levels of decision-making.
At the semantic level, the agent selects an information architecture---the overall layout, hierarchy, and spatial organization of the interface---that matches the functional constraints of the intent category to the structural properties of the return value (\cref{sec:principles}, P4).
At the component level, the agent selects and composes UI primitives to realize the chosen architecture.
Where a design system is provided, it supplies a library of standardized primitives with consistent styling, accessibility, and interaction behavior, constraining the component space without prescribing the layout; in its absence, the agent draws on its own knowledge of effective UI patterns.
Either way, semantic-level decisions---what the interface must accomplish---are made by the agent based on intent and data, while component-level decisions determine how those goals are realized.

\paragraph{Affordance generation.}
As part of the same construction step, the agent populates both channels of the affordance set.
Within $\Phi_0^s$, the agent generates \emph{view-embedded affordances}: interactive elements that arise naturally from the information architecture of $V_0$ itself.
These serve two complementary roles: some operate on data already visible---filters, sort controls, selection toggles---while others trigger new agent actions, such as booking, submitting, or expanding into further detail that has not yet been retrieved.
Both kinds are grounded in the current view context, but the latter initiates a new interaction cycle rather than reorganizing existing content.
Within $\Phi_0^{nl}$, the agent generates \emph{anticipatory affordances}: a dedicated interaction surface, separate from the main view, that surfaces intents the agent predicts to be likely next steps but that have no direct entry point within $V_0$.
Where view-embedded affordances produce parameterized events that the agent executes directly, anticipatory affordances dispatch as free-form intent signals that require agent interpretation---their cognitive advantage lies not in bypassing the natural language path, but in reducing the user's expressive burden from recall to recognition.
Together, the two sources implement the design goal stated in \cref{sec:principles}, P1: view-embedded affordances maximize the coverage of $\Phi_t^s$ for operations tightly coupled to the current information architecture, while anticipatory affordances ensure that the most valuable next steps beyond it remain discoverable without requiring the user to independently formulate them.

%% file: sections/4.system-architecture/4.3.App-Evolution.tex

\subsection{App Evolution}
\label{sec:system:evolution}

The defining characteristic of SaC---what distinguishes it from single-cycle generative UI---is that the application state persists and evolves across interaction cycles.
Each transition $s_t \rightarrow s_{t+1}$ carries forward the accumulated structure, data, and preference signals of all preceding cycles; the application does not reset, it builds.
This section describes how that transition is computed: how incoming events are interpreted, how the render function selects an update strategy, how affordances are updated, and how accumulated state is managed over time.

\paragraph{Event interpretation.}
Each interaction cycle begins with an incoming event $e$, originating from one of the two input channels.
When the user acts through a structured affordance $a \in \Phi_t^s$, the event carries explicit semantic content---the field being filtered, the item being selected, the action being triggered---reducing the user's expression cost relative to free-form natural language.
The agent still interprets this event in the context of $(V_t, C_t)$, though the degree of interpretation varies by implementation: affordances may be generated with execution parameters fully resolved at construction time, or resolved lazily at click time using the current context.
When the user acts through $\Phi_t^{nl}$---whether by typing freely or by selecting an anticipatory affordance---intent analysis is always required: the agent resolves the intent signal in the context of $(V_t, C_t)$, inferring what the user intends relative to what is already on screen and what prior cycles have established.
This context-sensitivity is what distinguishes evolution-time intent analysis from the cold-start case: the same utterance---"show me something cheaper"---carries different meaning depending on what $V_t$ currently displays and what $C_t$ records about prior refinements.

\paragraph{View update strategy selection.}
Once the event is interpreted, the agent determines an execution plan---interacting with $E$ through read or write operations as needed, performing computation, or reorganizing existing data---and then selects a view update strategy based on the structural compatibility between the result and the existing view $V_t$.
Three strategies are available, applied in order of increasing structural disruption.

The first is \emph{element-level update}: when the new data is structurally compatible with the current layout---additional items of the same type, updated values in an existing field, a refined subset of what is already displayed---the agent updates elements within the existing information architecture without altering the layout itself.
The user's spatial orientation is preserved; only the content within the structure changes.

The second is \emph{structural extension}: when the interaction opens a direction that is related to the current view but requires a different organization---a sub-topic, a comparative dimension, or a follow-up that the existing layout cannot accommodate---the agent appends a new section within the same application, typically as an additional tab or panel.
The existing structure is preserved and the new section is added alongside it, keeping the history of the interaction visible and navigable.

The third is \emph{application replacement}: when the user's intent diverges sufficiently from the current application's scope---a fundamentally different task, a context shift that renders the existing structure irrelevant---the agent initiates a new agentic application rather than extending the current one.
This is not a failure of the current application; it is a natural complexity boundary, consistent with the principle that accumulated structure should not be extended indefinitely when a fresh start better serves the user (\cref{sec:principles}, P6).

The selection among these three strategies is itself a design decision that intent analysis informs: a structured event from $\Phi_t^s$ typically signals convergence and favors element-level update, while a natural language input expressing an unanticipated direction more often triggers structural extension or replacement.

\paragraph{Affordance update.}
The render function produces not only a new view $V_{t+1}$ but also updated affordances across both channels.
Within $\Phi_{t+1}^s$, view-embedded affordances are updated as part of the new or updated information architecture: as the content of $V_{t+1}$ changes, the operations that are meaningful on that content change with it.
Within $\Phi_{t+1}^{nl}$, anticipatory affordances are recomputed based on the agent's updated model of the user's trajectory: which directions have already been explored, which adjacent intents now seem more likely given the interaction history, and which follow-up actions the current state makes newly relevant.
Over successive cycles, the anticipatory affordance set progressively specializes---surfacing fewer generic options and more task-specific ones---as the agent's model of this user's priorities sharpens.

\paragraph{State management.}
Each cycle appends new information to the agent context $C_t$: data and results from environment interaction, inferred preferences, interaction history, and the decisions made by the render function.
Over extended interactions, this accumulation poses a structural challenge: the very persistence that gives SaC its cognitive advantages can, if left unmanaged, produce a context too large to reason over effectively and an interface too complex to navigate (\cref{sec:principles}, P6).

The system manages this at two levels.
At the context level, interaction history is periodically compressed into summary representations that preserve key decisions and preference signals while discarding intermediate steps; raw history is retained where it may be needed for structural reference but is deprioritized in the agent's active reasoning window.
At the interface level, the three-strategy render model provides a natural complexity valve: structural extension adds sections rather than overloading a single layout, and application replacement resets complexity when the accumulated structure has outgrown its usefulness.

The specific thresholds and compression strategies for state management remain an open design question, and we discuss this further as a limitation in \cref{sec:discussion}.

%% file: sections/4.system-architecture/4.4.Quality-Assurance.tex

\subsection{Quality Assurance}
\label{sec:system:qa}

Quality assurance for agentic applications inherits from two distinct traditions.
From traditional software QA, it inherits the concern for functional correctness and interaction quality---whether the application does what it should, and whether users can effectively consume information and act through it.
From LLM output verification, it inherits the concern for content fidelity---whether the information the application presents is factually grounded, attributable, and sufficiently rich.
In conventional software and chat systems, these concerns are addressed separately; in SaC, all three must be satisfied simultaneously by a single generated artifact.
The additional challenge is that the artifact is dynamically generated: unlike traditional software QA, which validates against a fixed specification or product requirements document (PRD), SaC quality assurance must evaluate outputs that are structurally different every time---reference examples can serve as quality anchors, but there is no document to verify against.

\paragraph{Functional correctness.}
A generated agentic application must do what it implicitly promises to do.
This has two components.
The first is \emph{runnability}: the application executes without failure, all interactive elements are operable, and all referenced resources are accessible.
A broken affordance---a button that triggers no action, a data field that fails to load---is not merely an aesthetic failure; it breaks the interaction cycle and severs the $H \rightarrow A$ channel at the point where the user most expects it to function.
The second is \emph{content fidelity}: the degree to which the application's content is both accurate and sufficiently rich to serve the user's intent.
Accuracy requires that factual claims are grounded in retrievable sources, data values are precise rather than approximate, and the application does not present fabricated information as fact.
Richness requires that the content is informationally substantial---an application that is factually correct but sparse fails the same underlying requirement that the artifact faithfully serves the user's intent.
This component is structurally analogous to faithfulness evaluation in retrieval-augmented generation~\cite{lewis2020rag}: the agent's output should be entailed by the information retrieved from $E$, not confabulated beyond it, and should represent it with sufficient depth.

\paragraph{Interaction quality.}
Beyond correctness, a generated agentic application must support effective information consumption and action.
This also has two components.
The first is \emph{information architecture}: the structure of $V_t$ organizes content in a way that supports effective consumption---core information is visually prioritized, hierarchy matches semantic importance, and the layout guides the user through the content in a progression that reflects the logical structure of the task rather than presenting information as an undifferentiated flat list.
The second is \emph{visual quality}: the visual execution of $V_t$ meets a minimum design quality floor---text is legible, contrast ratios are sufficient, layout elements do not overflow or overlap, and the visual style is appropriate to the content domain.
These two components correspond to the UX and UI dimensions of traditional software evaluation---the distinction being that in SaC, both are properties of a generated artifact rather than a designed one, and must therefore be enforced through generative constraints rather than design review.

Whether these quality criteria are enforced inline---as constraints embedded in the generation and evolution prompts---or post-hoc, as a verification step after generation, is an engineering choice with a familiar tradeoff: inline enforcement guides the generative process from the start but risks prompt bloat and reduced generation precision, while post-hoc verification is modular and independently maintainable but requires regeneration on failure.
Both approaches are consistent with the quality criteria defined here; the appropriate balance is determined by system constraints rather than by the criteria themselves.

%% file: sections/4.system-architecture/4.5.Distribution-and-Consumption.tex

\subsection{Distribution and Reuse}
\label{sec:system:distribution}

Distribution in a SaC system is structurally different from distribution in a conventional content platform.
In a conventional platform, the distributed artifact is static: a fixed document, video, or page whose content does not change between creation and consumption.
In SaC, the artifact is a stateful object---$s_t = (V_t, \Phi_t^s, C_t)$---whose three components carry different kinds of information and serve different roles in the interaction lifecycle.
This distinction forces a more careful treatment of what it means to share, surface, and reuse an agentic application.

\paragraph{Two shareable objects.}
Because $s_t$ bundles generated interface structure, interaction affordances, and accumulated context data into a single object, sharing it wholesale is not always appropriate.
A user who has spent several interaction cycles exploring apartments in a specific city has produced a state $s_t$ whose $C_t$ is dense with personal constraints---budget range, preferred neighborhoods, commute requirements---even though the information architecture of $V_t$ and the affordance structure of $\Phi_t^s$ may be broadly useful to anyone looking for housing in the same city.
This motivates a distinction between two shareable objects.
\emph{State sharing} distributes the complete $s_t$, including its data---appropriate when the intent is to share a specific result, such as a curated apartment search or a travel itinerary built through exploration. 
\emph{Template sharing} distributes only the generative structure: the intent category, information architecture, and affordance design, stripped of user-specific context data. 
A recipient of a template does not receive a copy of the sharer's state; they receive the starting conditions for generating their own $s_0$, populated with their own data when the application is first invoked.
This is thus a mechanism for reusing the form of an agentic application without transferring its content.
Template sharing bears a surface resemblance to vibe coding~\cite{karpathy2025vibecoding}---the practice of generating application code from natural language descriptions---in that both produce a reusable generative artifact from a user's intent.
The distinction is conceptual: in vibe coding, the generated software is the goal---the template is the final product and a deliverable artifact.
In SaC, the template is a starting point with established architecture, consistent with the framing in \cref{sec:lifecycle} that SaC treats software as a medium to communicate with the agent.

\paragraph{Data-driven update.}
A further distinction arises when the data underlying an agentic application changes after distribution.
Consider a currency conversion application shared via state sharing: the exchange rates embedded in $C_t$ at the time of sharing may be stale by the time a recipient opens it.
This is not an interaction-driven state transition---no user action has been taken---but a \emph{data-driven update}: the underlying environment $E$ has changed, and the application's content should reflect that change.
The distinction matters for versioning: interaction-driven transitions produce new states in the application's lifecycle and are part of its evolution history; data-driven update refreshes the content of an existing state without advancing the lifecycle.
A currency application that updates its rates on open has not evolved---it has updated.
Template sharing handles this naturally, since each invocation re-fetches data from $E$ at the time of use; state sharing requires an explicit decision about whether embedded data is treated as a static snapshot or as a live reference to be updated on access.

%% file: sections/5.evaluation/5.1.Methodology.tex
\subsection{Methodology}
Our evaluation pursues two goals: to demonstrate that the \emph{Software as Content} paradigm is technically realizable,
and to characterize the expressive range of agentic applications across diverse task types.
We evaluate through a reference implementation---a playground environment supporting app generation and evolution via element-level update and structural extension.

Rather than a large-scale user study, we adopt a scenario-based approach~\cite{carroll2000making}:
three walkthrough scenarios spanning the selection, exploration, and execution intent categories from the taxonomy in Section~\ref{sec:taxonomy}, followed by two boundary cases that characterize conditions under which structured interaction is not the appropriate modality.
Each scenario is presented as an interaction trace that illustrates how the formal model instantiates in practice--- demonstrating that agentic applications can be generated and evolved without failure, and that the resulting evolution trajectories exhibit the bidirectional, state-accumulating properties characterized in Section~\ref{sec:framework}.
A more comprehensive empirical evaluation, including comparative user studies and quantitative metrics, is left for future work as the open-source implementation matures.

%% file: sections/5.evaluation/5.2.Scenario-Walkthroughs.tex
\subsection{Scenario Walkthroughs}
\label{sec:scenarios}

The following four scenarios span three intent categories from the taxonomy in Section~\ref{sec:taxonomy}---selection, exploration, and execution---each drawn from authentic use cases rather than constructed examples.
The three scenarios are sequenced to demonstrate qualitatively distinct properties of the paradigm: the first foregrounds information synthesis and affordance anticipation under selection intent; the second demonstrates evolution across multiple cycles and modalities under exploration intent; the third illustrates context-sensitive execution scaffolding.

\subsubsection{Multi-Constraint Selection: Car Rental Planning}
\label{sec:scenario-carrental}

\begin{figure*}[t]
    \centering
    \includegraphics[width=\textwidth]{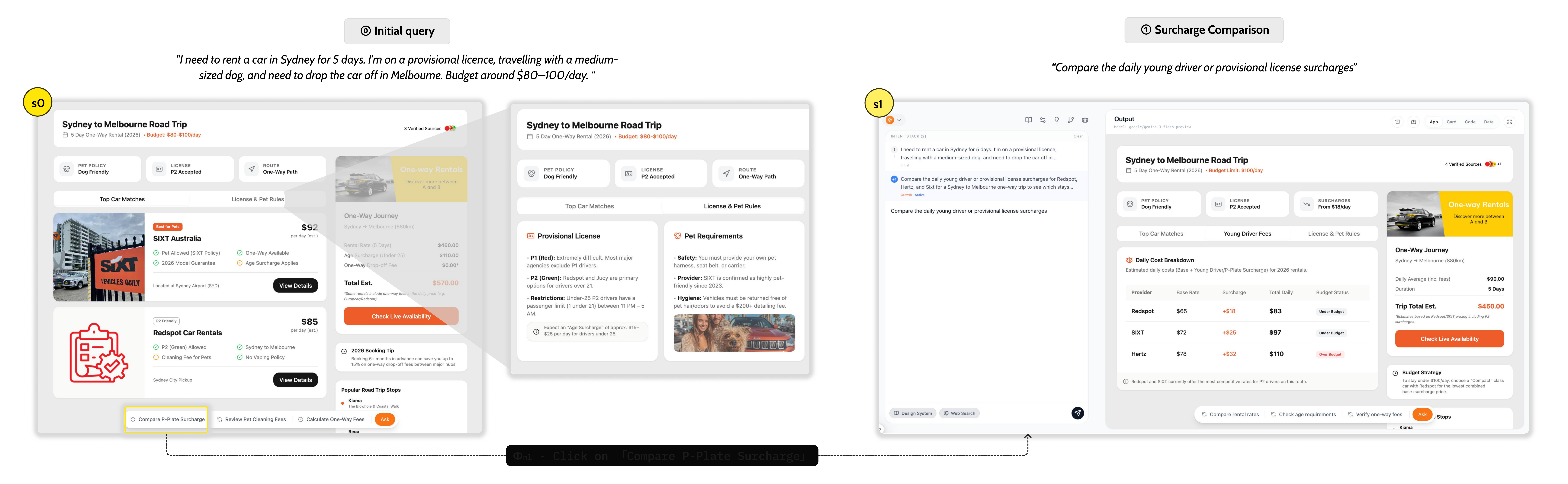}
    \caption{Scenario 1: Car rental planning scenario. \textit{Cold start} ($s_0$, left): initial application state generated from a query with three interacting constraints---provisional licence, dog, and one-way drop-off. \textit{Evolution} ($s_0 \to s_1$, right): structural extension triggered by the anticipatory affordance \textit{Compare P-Plate Surcharge} ($\Phi^{nl}$), appending a \textit{Young Driver Fees} tab and a provider cost comparison table.}
    \label{fig:carrental}
\end{figure*}

As shown in \ref{fig:carrental}, a user queries the system with three interacting constraints: a provisional (P2) licence, a medium-sized dog, and a one-way Sydney-to-Melbourne drop-off, with a daily budget of \$80--100.
Notably, no rental platform exposes provisional licence eligibility as a filterable attribute; the agent retrieves this information by synthesizing individual company policy documents across providers and resolving compatibility against the user's stated licence type---a cross-source synthesis task that would require the user to manually visit multiple websites in the absence of an agent.

The generated $s_0$ organises these constraints into a decision architecture: three constraint badges (Dog Friendly / P2 Accepted / One-Way Path) serve as a persistent constraint summary; two tabs separate candidate matches from licence and pet rules; and a cost breakdown panel aggregates the financial implications of all three constraints into a single estimated total (Figure~\ref{fig:carrental}, left).
The anticipatory affordance set $\Phi^{nl}_0$ surfaces three predicted follow-up intents---\textit{Compare P-Plate Surcharge}, \textit{Review Pet Cleaning Fees}, and \textit{Calculate One-Way Fees}---none of which require the user to articulate them in natural language.

The user selects \textit{Compare P-Plate Surcharge}, triggering a structural extension: a new \textit{Young Driver Fees} tab is appended to the existing architecture without displacing the prior structure, and the updated view introduces a cost comparison table across three providers with a \textit{Budget Status} column that persists the \$80--100/day constraint from $s_0$ into the evolved state $s_1$ as an explicit decision signal (Figure~\ref{fig:carrental}, right).
The constraint summary updates to reflect the new focus---\textit{Surcharges From \$18/day} replaces the route badge---while Dog Friendly and P2 Accepted are preserved, demonstrating context-sensitive evolution rather than wholesale regeneration.

This scenario illustrates the primary value of SaC under selection intent: the agent's ability to synthesise information from heterogeneous sources into a coherent decision architecture at $s_0$, and an affordance set that anticipates the user's most probable next decision without requiring natural language reformulation.

\subsubsection{Exploration: Weekend BBQ Planning}
\label{sec:scenario-bbq}

\begin{figure*}[t]
    \centering
    \includegraphics[width=\textwidth]{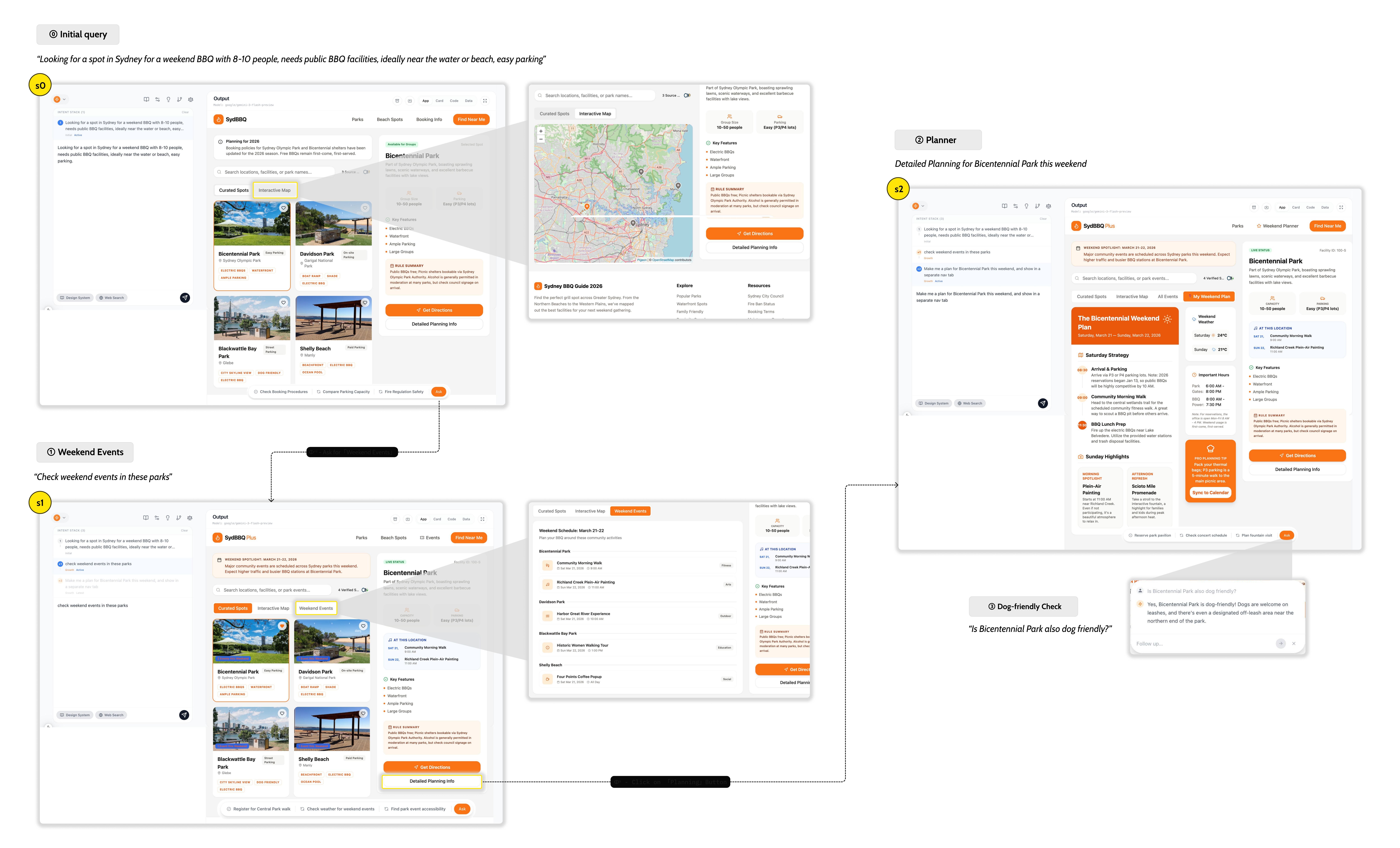}
    \caption{Scenario 2: Weekend BBQ planning scenario. \textit{Cold start} ($s_0$): dual-view architecture with \textit{Curated Spots} and \textit{Interactive Map}. \textit{First evolution} ($s_0 \to s_1$): structural extension via $\Phi^{nl}$, adding \textit{Weekend Events} tab with in-place card updates. \textit{Second evolution} ($s_1 \to s_2$): structural extension via $\Phi^s$, converging to a location-specific weekend plan. \textit{Plain text reply}: factual query answered directly without updating $s_2$.}
    \label{fig:bbq}
\end{figure*}

As illustrated in \ref{fig:bbq}, a user queries the system for a public BBQ spot in Sydney suitable for 8--10 people, near water, with easy parking.
Unlike selection intent, where the user arrives with a defined decision criterion, this task is exploratory: the user has a rough preference but no precise goal, and the interaction is expected to progressively narrow toward a decision rather than filter toward one.

The generated $s_0$ reflects the open-ended nature of exploration intent through a dual-view architecture.
The \textit{Curated Spots} view presents locations as image-rich cards with facility tags and a persistent detail panel, supporting attribute-based browsing without committing to a decision.
The \textit{Interactive Map} view renders the same options as georeferenced markers across Sydney, making spatial relationships between options directly perceptible rather than requiring the user to mentally reconstruct them from prose.
Switching between the two views is a view-embedded affordance triggering an element-level update, allowing the user to shift between attribute reasoning and spatial reasoning as their exploration evolves.

As the user browses, they submit a natural language query through $\Phi^{nl}$: \textit{``check weekend events in these parks.''}
The agent extends the application with a new \textit{Weekend Events} tab via structural extension ($s_0 \to s_1$), preserving the existing \textit{Curated Spots} and \textit{Interactive Map} views while appending a new facet of information relevant to the user's emerging intent.
Notably, the evolution is not limited to the new tab: existing card elements are updated in place to surface event-relevant signals---an \textit{Events this Weekend} badge appears on applicable location cards, and the detail panel gains an \textit{At This Location} section listing scheduled events---without disrupting the overall layout.
The application accumulates structure rather than resetting, keeping the full history of the exploration navigable.

The user's interest converges on Bicentennial Park.
They select the \textit{Planning} affordance ($\Phi^s$), triggering a second structural extension ($s_1 \to s_2$) that generates a full weekend plan for the selected location---arrival strategy, schedule, weather, and logistics---within the same application.
The transition from open-ended exploration to task-specific planning is mediated entirely through the application's evolving structure, without requiring the user to restart or reframe their query.

A final natural language query---\textit{``is Bicentennial Park also dog friendly?''}---is answered by the agent as a plain text reply without updating $s_2$, consistent with Principle~2 (match output modality to task structure).
The application remains stable, preserving the user's planning context across the exchange.

The full trajectory---from open-ended browsing, through event-driven structural extension, to location-specific planning, with a plain text reply interspersed---illustrates how a single agentic application can sustain an exploratory task across qualitatively different interaction modes without resetting its accumulated context.

\subsubsection{Personalised Execution Guidance: SSN Application for F-1 OPT Students}
\label{sec:scenario-ssn}

\begin{figure*}[t]
    \centering
    \includegraphics[width=\textwidth]{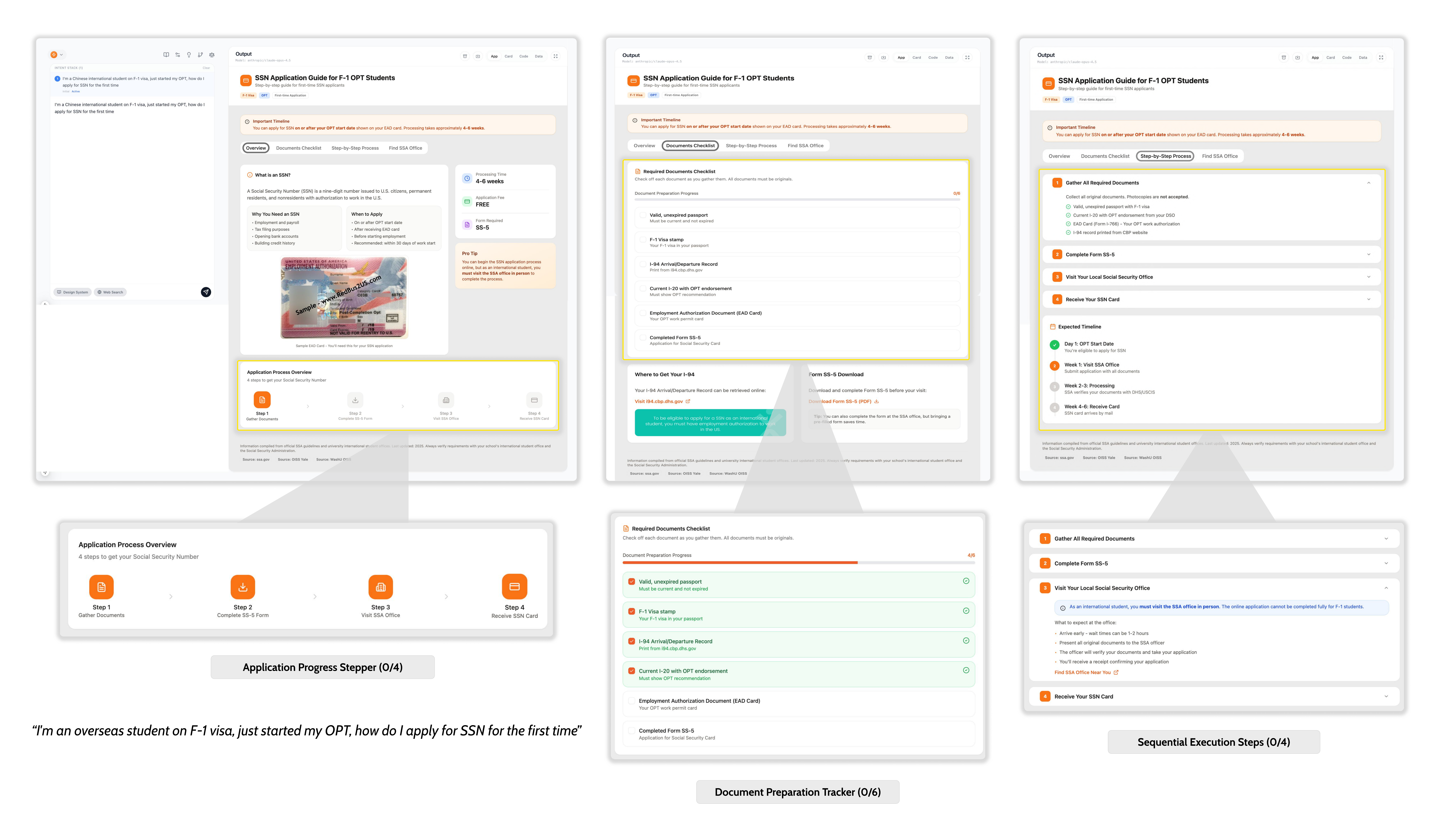}
    \caption{Scenario 3: SSN application guidance scenario. The generated $s_0$ organises a personalised execution path into four tabs based on the user's F-1 OPT context. Three tabs are shown: \textit{Overview} with an application progress stepper; \textit{Documents Checklist} with a preparation tracker specific to the F-1 OPT document combination; and \textit{Step-by-Step Process} with a sequential execution layout and OPT-grounded timeline. Tab navigation triggers element-level updates within a stable application architecture.}
    \label{fig:ssn}
\end{figure*}

As demonstrated in \ref{fig:ssn}, a user queries the system as a Chinese international student on an F-1 visa who has just started OPT and needs to apply for an SSN for the first time.
The query carries sufficient personal context for the agent to determine not only what information to retrieve, but what execution path applies to this specific user---a path that differs substantially from the standard SSN application process for U.S. citizens or other visa categories.

The generated $s_0$ organises the application into a four-tab architecture, each tab corresponding to a distinct phase of the execution workflow (Figure~\ref{fig:ssn}).
The \textit{Overview} tab surfaces an \textit{Application Progress} stepper that externalises the user's position within the overall process---a persistent representation of execution state that chat's append-only medium cannot maintain across turns.
The \textit{Documents Checklist} tab presents a \textit{Document Preparation Progress} tracker (0/6) with a checklist specific to the F-1 OPT combination: passport with F-1 visa stamp, I-20 with OPT endorsement, EAD card, and I-94 record---documents that a generic SSN guide would not surface without knowing the user's visa status.
The \textit{Step-by-Step Process} tab renders the four execution steps as an expandable sequential layout, with an \textit{Expected Timeline} grounded in the user's OPT start date as the reference point.

Across all three tabs, the structured affordances $\Phi^s$---\textit{Download Form SS-5}, \textit{Find SSA Office Near Me}, \textit{Start Application}---function as direct channels through which the user can trigger agent actions at each stage of the workflow.
In the SaC paradigm, these affordances are not hyperlinks to external resources; they are parameterised entry points into agent execution, carrying the user's context forward into each subsequent action without requiring reformulation.

The result is an execution scaffold that is both personalised and persistent: the workflow, documents, and timeline are specific to this user's visa status, and the application externalises execution state in a form that survives across turns---something chat's append-only medium structurally cannot provide.

%% file: sections/5.evaluation/5.3.Boundaries-of-the-SaC-Scope.tex
\subsection{Boundaries of the SaC Scope}
\label{sec:boundary}

The preceding scenarios demonstrate SaC's expressive range across selection, exploration, and execution intent.
Characterising where SaC is \emph{not} the appropriate modality is equally important---not as an admission of failure, but as a necessary part of understanding the paradigm.
We identify two structurally distinct boundary conditions, each arising from a different property of the interaction rather than a limitation of the implementation.

\subsubsection{Socio-Emotional Interaction}
\label{sec:boundary-emotional}

\begin{figure}[t]
    \centering
    \includegraphics[width=\columnwidth]{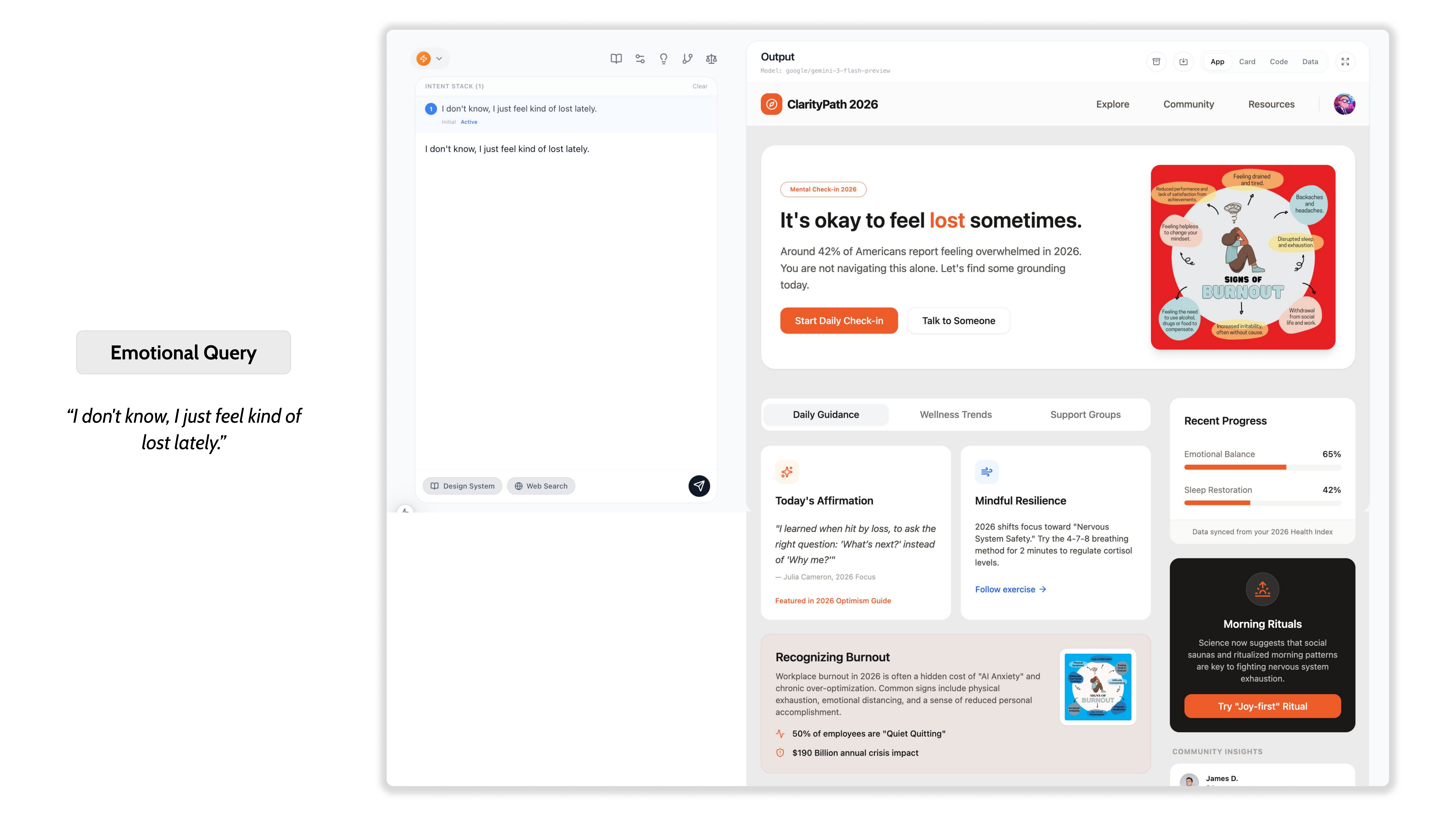}
    \caption{Socio-emotional boundary case. The query \textit{``I don't know, I just feel kind of lost lately''} generates a wellness dashboard with quantified emotional metrics---\textit{Emotional Balance: 65\%}, \textit{Sleep Restoration: 42\%}. 
    The mismatch is not in the content but in the ontology: a structured interface converts an emotional expression into a measurable state to be resolved, reframing the user as a case rather than an interlocutor.}
    \label{fig:boundary-emotional}
\end{figure}

Consider a user who submits the query \textit{``I don't know, I just feel kind of lost lately.''}
As shown in Figure~\ref{fig:boundary-emotional}, the system generates a wellness dashboard---a structured interface with burnout indicators, daily guidance tabs, and a \textit{Recent Progress} panel reporting \textit{Emotional Balance: 65\%}.
The application is informationally coherent, but it mismatches the nature of the interaction.
The user's query does not express a problem to be diagnosed; it expresses a state to be received.

This distinction has a clinical parallel. 
In therapeutic practice, clinicians follow an open-to-closed questioning discipline: beginning with ``how have you been feeling lately?'' rather than ``rate your anxiety from one to ten.''
The structured question is more efficient and yields more measurable data. 
But starting with it changes the nature of the exchange---it signals that the clinician's role is to evaluate and categorise, not to listen and understand.
The structure itself, before any content is exchanged, reframes what kind of relationship is taking place.

The wellness dashboard in Figure~\ref{fig:boundary-emotional} performs the same reframing.
Structuring the query into an agentic application performs an implicit ontological conversion: it treats an open emotional expression as a solvable task with measurable dimensions, and positions the system as an evaluator rather than an interlocutor.
The \textit{Emotional Balance: 65\%} metric is the clearest symptom of this conversion---not because the metric is inaccurate, but because quantifying the user's inner state as a progress indicator fundamentally reframes what kind of exchange is taking place.

In socio-emotional interaction, the medium itself carries meaning.
A conversational text exchange signals a relationship of mutual presence---two parties speaking to each other.
A structured interface signals an asymmetric relationship---one party being assessed by a system.
This asymmetry is not a matter of content or visual design; it is constitutive of the interaction's social character.
For this class of interaction, the more appropriate response would not be a lighter or simpler agentic application---it would be the natural language channel alone, without any instantiation of structured UI.

\subsubsection{Open-Ended Conceptual Exploration}
\label{sec:boundary-conceptual}
\begin{figure}[t]
    \centering
    \includegraphics[width=\columnwidth]{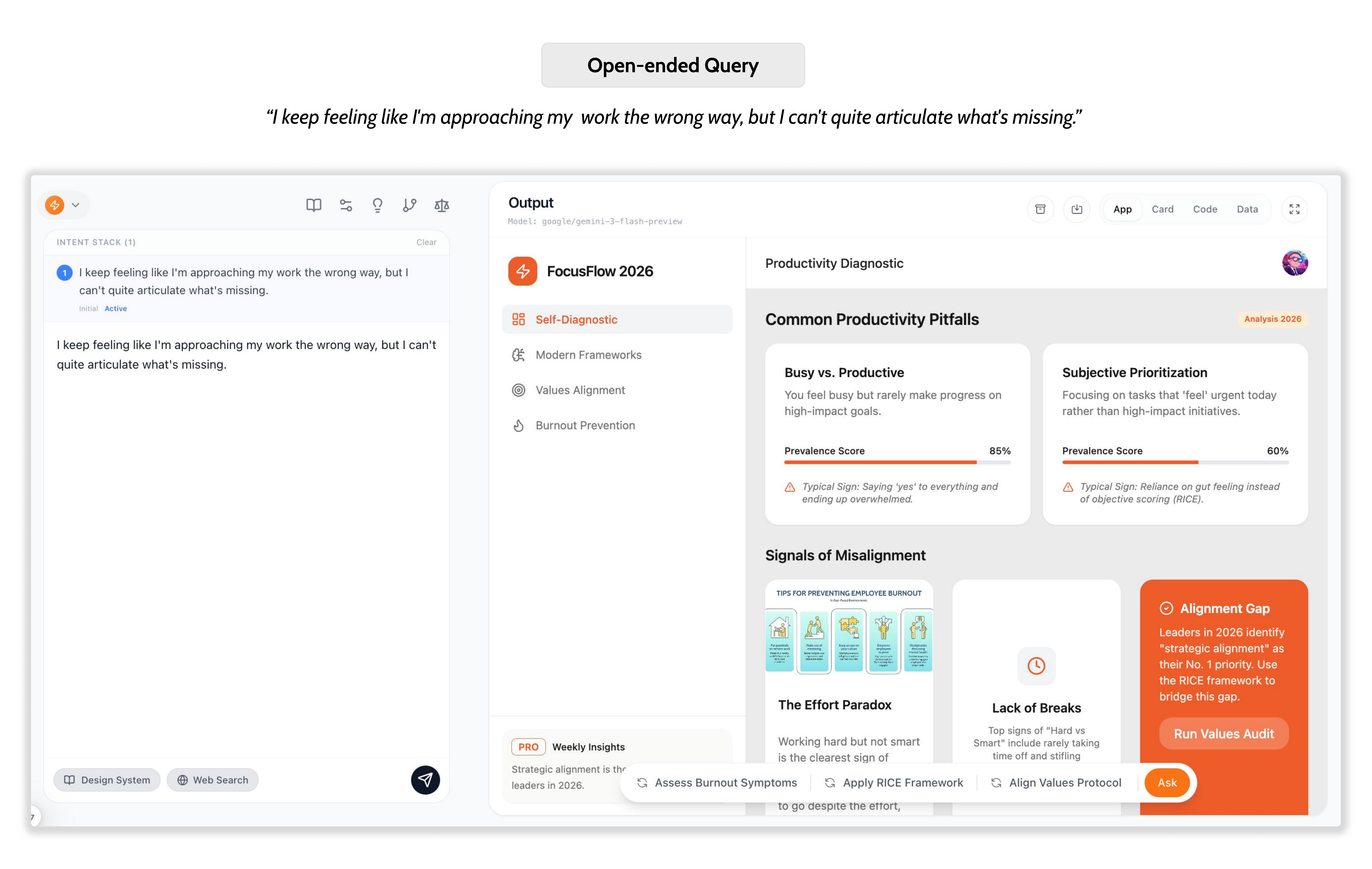}
    \caption{Open-ended conceptual exploration boundary case.
    The query \textit{``I keep feeling like I'm approaching my work the wrong way, but I can't quite articulate what's missing''} generates a productivity diagnostic dashboard with prevalence scores, framework tabs, and affordances such as \textit{Apply RICE Framework} and \textit{Run Values Audit}.
    The application is not incoherent; it is premature---the agent has committed to a narrowed problem frame before the user has found one.}
    \label{fig:boundary-conceptual}
\end{figure}

The preceding boundary arose from the social character of the interaction.
Another situation arises from its epistemic character, and applies even when the query carries no emotional valence.

Consider a user who asks: \textit{``I keep feeling like I'm approaching my work the wrong way, but I can't quite articulate what's missing.''}
As shown in Figure\ref{fig:boundary-conceptual}, a structured response---tabs organising possible frameworks, affordances for selecting relevant dimensions---is not obviously inappropriate here the way a wellness dashboard was in the previous case.
The query is cognitive rather than emotional, and SaC routinely handles exploratory intent.
However, there is a specific condition under which structure becomes a liability: when the user does not yet know what dimensions are relevant, and the value of the interaction lies precisely in discovering them.

Generating an agentic application requires the agent to commit, at construction time, to a set of dimensions worth presenting and a set of affordances worth exposing.
For tasks where the problem space is already legible---even partially---this commitment is productive: it organises what is known and makes the next steps visible.
But when the problem space itself is what the user is trying to discover, the agent's structural commitment is premature.
The interface does not reflect the user's understanding of the problem; it substitutes the agent's understanding for it.
What the user needed was space to think out loud; what they received was a pre-organised view of what the agent thinks they should be thinking about.

This is a narrower condition than it might appear.
SaC's exploration intent category is designed precisely to support open-ended inquiry---progressive disclosure, bidirectional refinement, and divergence handling are all part of its design.
The boundary applies not to exploration in general, but to the specific case where the user's intent is genuinely pre-structural: where they do not yet have a question, only a feeling that a question exists.
In this case, the natural language channel---unhurried, unstructured, without anticipatory affordances---is the more appropriate medium, because it does not force the interaction into a shape before that shape has been found.

Together, the scenarios and boundary cases in this section establish two complementary findings: that agentic applications can be reliably generated and evolved across diverse task types, and that the conditions under which structured interaction is appropriate are themselves a meaningful object of study---one that points toward a broader research agenda we take up in the following section.

%% file: sections/6.discussion/6.1.Rethinking-the-Role-of-the-Interface-in-Agent-Systems.tex
\subsection{Rethinking the Role of the Interface in Agent Systems}
\label{sec:rethink}

SaC challenges a chat paradigm-inherited assumption at its foundation: interaction quality is a downstream function of agent capability, and as agents become more powerful, the quality of human–agent collaboration will improve accordingly. 
Capability and interaction architecture are orthogonal dimensions.
The structural constraints of chat identified in Section~\ref{sec:intro}—representation mismatch, interaction entropy, ephemeral state—are not resolved by making the agent smarter;
they are properties of the channel, not of the reasoning that flows through it.
A more capable agent operating through a chat interface remains constrained by that interface's architecture.
Conversely, the same agent operating through an agentic application resolves all three constraints without any change to its underlying capabilities.
This orthogonality has a direct implication for how the field evaluates progress:
measuring agent capability in isolation, without accounting for the interaction layer through which that capability is exercised, captures only half of what determines collaborative outcome.

The deeper reason for this orthogonality lies in what agents actually are in relation to users.
The dominant framing positions the agent as a conversational partner—an entity the user speaks to directly.
But this framing obscures a more accurate structural analogy.
Traditional software consists of a pre-built frontend and a pre-built backend;
the user interacts with the frontend, which translates their actions into operations on the backend.
SaC completes this analogy at both ends: the frontend is generated according to the user's intent and the data and context at hand,
and the backend is the agent itself.
The user is not speaking to the agent; they are using a system in which the agent is the engine.
This is not a superficial reframing—it has consequences for how we think about what the interface is for.
Users do not interact with engines; they interact with the systems that engines power.
The agentic application is that integrated system, wielding the agent as a component while providing a complete service to the user.

This framing directly addresses a recurring claim in agent-oriented discourse: that as natural language becomes the dominant input modality, graphical interfaces will become unnecessary.
The claim conflates two distinct things.
Natural language as an input channel is itself a form of interface—and a particularly low-density one, as we argued in Section~\ref{sec:intro}.
The question is never whether an interface exists, but what kind.
More fundamentally, the claim gets the direction of the relationship backwards.
A more powerful engine does not reduce the importance of the interface that exposes it; it increases it.
The richer the agent's action space, the more the user needs a well-structured interface to navigate that space—to understand what is available, to direct execution with precision, and to observe the results without cognitive overload.
The analogy is not subtle: a more powerful engine does not make the dashboard less important.
It makes it more important.
As agent capabilities continue to expand, the interaction layer is not a diminishing concern—it is an increasingly critical one, and one that SaC positions as a first-class research object rather than an engineering afterthought.

%% file: sections/6.discussion/6.2.Limitation.tex
\subsection{Limitations}
\label{sec:limitations}

Section~\ref{sec:evaluation} identifies the principled boundaries of SaC's scope:
socio-emotional interaction and pre-structural conceptual exploration are two classes of user intent for which unstructured chat is the more appropriate response form.
What the evaluation does not resolve is the prior question: how does a system reliably determine, before generation begins, whether a given interaction calls for a structured agentic application or for the natural language channel alone?
The signals are subtle—a query that superficially resembles an exploration intent may carry an emotional or epistemic character that only becomes apparent in hindsight.
This judgment is fundamentally empirical: the patterns that distinguish contexts where SaC is the right form from those where it is not will likely only converge through sustained exposure to real user interactions at scale, as edge cases accumulate and the boundary conditions become legible from data rather than from theory.

The current reference implementation has two constraints worth distinguishing by their nature.
The first is latency: at present, initial application generation takes on the order of thirty seconds, which places a perceptible burden on the interaction.
This is an implementation constraint, not a paradigm constraint—and the trajectory of inference hardware makes it a temporary one.
Taalas, a hardware startup that physically etches model weights into custom silicon, has demonstrated inference throughput of 16,000 to 17,000 tokens per second per user on its HC1 chip~\cite{taalas2026hc1}—roughly two orders of magnitude beyond current GPU baselines.
As specialised inference hardware of this kind enters broader deployment, the latency profile of SaC generation will improve correspondingly, and the interaction experience at that speed will be qualitatively different from what the current implementation can demonstrate.

The second constraint is of a different character.
The current implementation does not perfectly support write-capable agent execution—the agent can retrieve and synthesise information from the environment, but cannot act on it: submitting forms, writing data, triggering external services, or modifying state in external systems.
This is not a gap that inference speed or prompt engineering will close.
Write-capable execution requires solving a set of interconnected backend problems: reliable agent control over external systems with heterogeneous APIs, transactional consistency when multi-step workflows partially fail, and frontend state synchronisation when agent-initiated writes change the ground truth that the agentic application displays.
At a deeper level, this is a backend architecture problem for dynamic software as a class—not specific to SaC's interaction model, but a necessary condition for SaC to realise its full execution intent category.
It requires systematic technical exploration rather than incremental refinement of the current pipeline.

The design of the human-facing interaction layer also remains open.
The current implementation surfaces the natural language channel as an intent bar positioned alongside the agentic application—one concrete answer to the question of how $\Phi^{nl}_t$ and $\Phi^s_t$ coexist in a single interface.
Whether this is the ultimate solution is not yet known.
How users naturally switch between structured affordances and free-form input, where they expect the language channel to appear, and how its presence should change as the application evolves—these are product design questions that require sustained iteration against real user behaviour, and the current design should be understood as a starting point rather than a settled solution.

Finally, the evaluation presented in Section~\ref{sec:evaluation} is an existence proof, not an empirical characterisation.
It demonstrates that SaC is technically realisable and expressively coherent across diverse task types; it does not quantify how SaC compares to chat-based alternatives in controlled conditions, how interaction quality evolves over extended sessions, or how different user populations engage with agentic applications.
User studies and longitudinal evaluation are the natural next step.
But the more fundamental gap is earlier in the research pipeline: there is currently no established benchmark for evaluating agentic applications as an interaction modality.
What constitutes a well-formed agentic application?
How should interaction expressiveness be measured across evolution cycles?
What metrics capture the quality of the affordance set $\Phi^s_t$ relative to the user's actual intent trajectory?
These questions do not yet have agreed answers, and designing the evaluation framework is itself an open research problem—one that SaC, as a first concrete instantiation of this paradigm, is now in a position to motivate.

%% file: sections/6.discussion/6.3.Future-Work.tex
\subsection{Future Work}
\label{sec:futurework}

The most immediate direction is empirical.
The evaluation presented in this paper demonstrates technical feasibility and expressive range;
it does not yet characterise how real users engage with agentic applications over time, how interaction quality evolves across sessions, or how SaC compares to chat-based alternatives under controlled conditions.
User studies across diverse task domains and longitudinal evaluation across extended interactions are the necessary next step.
Closely related is the scope fitness problem identified in Section~\ref{sec:limitations}: as real interaction data accumulates, it becomes possible to study the signals that reliably precede appropriate and inappropriate contexts for SaC, and to build more principled guidance into the modality assessment step of the pipeline.

A second direction concerns evaluation infrastructure.
There is currently no benchmark for assessing agentic applications as an interaction modality—no agreed metrics for interaction expressiveness, affordance quality, or the coherence of application evolution across cycles.
Designing this evaluation framework is itself an open research problem, and one that requires empirical grounding in real user behaviour rather than purely theoretical specification.
SaC, as a first concrete instantiation of the dynamic software paradigm, is now in a position to motivate and anchor this line of work.

Further ahead, the distribution and reuse mechanisms outlined in Section~\ref{sec:system:distribution} point toward a natural extension: an ecosystem in which agentic applications are shared, discovered, and collectively iterated upon.
The distinction between state sharing and template sharing introduced in this paper provides a starting point, but the design of a marketplace or content platform for agentic applications raises its own questions around provenance, versioning, and the social dynamics of sharing artifacts whose value is inseparable from the interaction history that produced them.

Finally, the SaC interaction model points toward a standardisation opportunity.
Just as the Model Context Protocol standardised the interface between agents and external tools~\cite{anthropic2024mcp}, a SaC protocol could standardise the interface between agents and the interaction layer—specifying how agents communicate application state, how frontends render and evolve agentic applications, and how structured affordances encode and dispatch user intent.
Such a protocol would decouple the interaction layer from any particular agent implementation, enabling the ecosystem of agentic applications to develop independently of the underlying agent infrastructure.
This remains a longer-term direction, but one whose foundations are already visible in the architecture described in Section~\ref{sec:system}.

%% file: sections/7.conclusion.tex
This paper introduced Software as Content, a paradigm that reconceives the human–agent interaction layer as a dynamically generated, evolving agentic application—a shared medium through which human intent and agent execution are mutually expressed, and through which information is organised, actions are surfaced, and collaborative outcomes emerge.
We formalised the interaction cycle and lifecycle governing this medium;
proposed a system architecture and design principles for its generation and evolution;
and demonstrated through a reference implementation and scenario-based evaluation that the paradigm is technically realisable and expressively coherent across diverse task domains.

The deeper contribution is what SaC makes newly possible to study.
Dynamic software—software that is generated on demand, evolves through use, and is discarded when its purpose is served—has long been a conceptual possibility, but has lacked a sufficiently realised instance to serve as a research object.
Questions about how such software should be designed, evaluated, and distributed; how users engage with interfaces that did not exist before they asked for them; and what it means for a piece of software to be consumed rather than installed—these have remained largely unaddressed, not for lack of relevance, but because no sufficiently realized instance of dynamic software existed to study.
SaC is an attempt to provide that foundation.